\documentclass[twocolumn,trackchanges]{aastex7}

\usepackage{threeparttable}
\usepackage{amsmath}

\begin{document}

\title{SED Fitting of Globular Clusters in NGC 4874:  Masses and Metallicities}

\author[orcid=0000-0001-5290-6275,sname='Hartman']{Kate Hartman}
\affiliation{McMaster University, Department of Physics \& Astronomy}
\email[show]{hartmk2@mcmaster.ca}  

\author[orcid=0000-0001-8762-5772,sname='Harris']{William E. Harris} 
\affiliation{McMaster University, Department of Physics \& Astronomy}
\email{harrisw@mcmaster.ca}

\author[orcid=0009-0002-8949-4547,sname='Kim']{Jinoo Kim}
\affiliation{McMaster University, Department of Physics \& Astronomy}
\email{kim697@mcmaster.ca}

\begin{abstract}

In most nearby galaxies, photometry of the integrated light of their globular clusters (GCs) has been obtained in only two filters, yielding just a single color index.  However, NGC 4874, the brightest central galaxy in the Coma cluster, now has Hubble Space Telescope (HST) photometry available in ten filters, giving us a special opportunity to test SED fitting procedures on GCs in distant galaxies.  We fitted 29 of the brightest GCs with a library of SEDs from E-MILES and calculated the best-fit metallicity and mass of each cluster.  Using the fitted masses and luminosities derived from the reddest magnitudes, in the flat portion of the GC spectrum, we also calculated inferred mass-to-light ratios for our sample GCs; these were in the range (M/L) $\simeq 2 - 4$, slightly larger than the average values for Milky Way GCs but within the conventional range.

\end{abstract}



\section{Introduction} \label{sec:intro}

The rapidly growing James Webb Space Telescope (JWST) archive includes a myriad of images suitable for the study of globular clusters (GCs) in cosmologically  distant galaxies.  The ability of JWST/NIRCAM to image in two filters at once, along with its large suite of filters, puts multi-filter photometry and all of the analysis techniques it enables well within reach for GC studies \citep{harris2023jwst,harris2024jwst}.  \cite{faisst2022tinythings}, for example, performed SED (spectral energy distribution) fitting on young, compact star clusters in the SMACS J0723.3-7327 galaxy cluster using JWST photometry; given the galaxy cluster's redshift, the authors fitted for star cluster metallicity, mass, and age simultaneously, finding best-fit metallicity and mass values in the ranges expected for present-day globular clusters, but much younger ages.

Compared to SED fitting to the integrated light of entire galaxies with their extended and complex star formation histories, SED fitting of GCs is relatively simple.  GCs are old, compact, gravitationally bound systems of stars that formed in the early universe and have survived to the present day.  They appear as unresolved point sources in all but the nearest of galaxies, allowing for efficient photometric observations, but unlike younger clusters with rapidly evolving massive stars, the SEDs of old clusters change relatively slowly as they age.  This allows us to model the integrated light from GCs reasonably well with single-age simple stellar population (SSP) models.  Additionally, the distance (redshift) of host galaxies with observable GC systems tends to be well known, allowing us to fit other properties, such as cluster metallicity and mass, with more precision.  Almost all GC masses in previous literature are estimated based on luminosity and an assumed M/L derived from Milky Way GCs, typically no greater than $(M/L) = 2$ (\citet{fan2018m31sed}, for example, derive GC masses from several M/L sources), despite the relationship between GC mass and therefore M/L and host galaxy mass \citep{villegas2010gclum,harris2017masstolight}.  For distant systems where direct spectroscopy is not feasible, SED fitting can provide a way to estimate these cluster properties through multi-band photometry that covers a wide range of wavelengths, reducing the assumptions about fundamental cluster characteristics and reliance on luminosities in just one or two bandpasses.

Typically, GCs in nearby galaxies have almost always been observed in just two filters to produce a single photometric color index.  Hundreds of GC systems have been measured this way \citep[e.g.][among many others]{peng2006nonlinear,villegas2010gclum,harris2013stellarmass,harris2023colours}; a rare exception is the multifilter data for the GCs in M31 \citep{fan2016m31sed,fan2018m31sed}.  However, JWST programs such as PEARLS \citep{windhorst2023pearls}, CANUCS \citep{willott2022canucs}, UNCOVER \citep{bezanson2024uncover}, and VENUS \citep[][GO-6882]{fujimoto2025venus}, involve multi-band photometry of lensing-cluster targets at redshifts typically in the range  $z \simeq 0.2 - 0.5$, in which the major galaxies have large numbers of GCs.  As a low-redshift complement to these programs, new photometric data available for the Coma cluster central giant NGC 4874, to be described below, gives us an unusual opportunity to test SED fitting procedures that will become increasingly widely used with higher redshift JWST data.  While there is currently no data covering both HST's blue bands and JWST's red bands, JWST's NIRCam includes filters falling to both the blue and red sides of the GC SED peak, providing leverage on both metallicity and mass; techniques for fitting metallicity and mass using HST data will therefore be directly transferrable to JWST data.

We present a proof-of-concept study on the viability of SED fitting for low-redshift GC systems (GCSs).  In Section \ref{sec:data}, we outline our photometric testing data and the SEDs used in our fitting routine.  Section \ref{sec:fitting} provides details on the calculations, mechanics, and outputs of the routine.  We present fitting results and other derived quantities in Section \ref{sec:results} and compare them to observations.  Finally, in Section \ref{sec:conclusions}, we discuss possible next steps and followup opportunities.

\section{Data} \label{sec:data}

\subsection{Photometry} \label{subsec:photometry}

\begin{figure}
    \centering
    \includegraphics[width=0.47\textwidth]{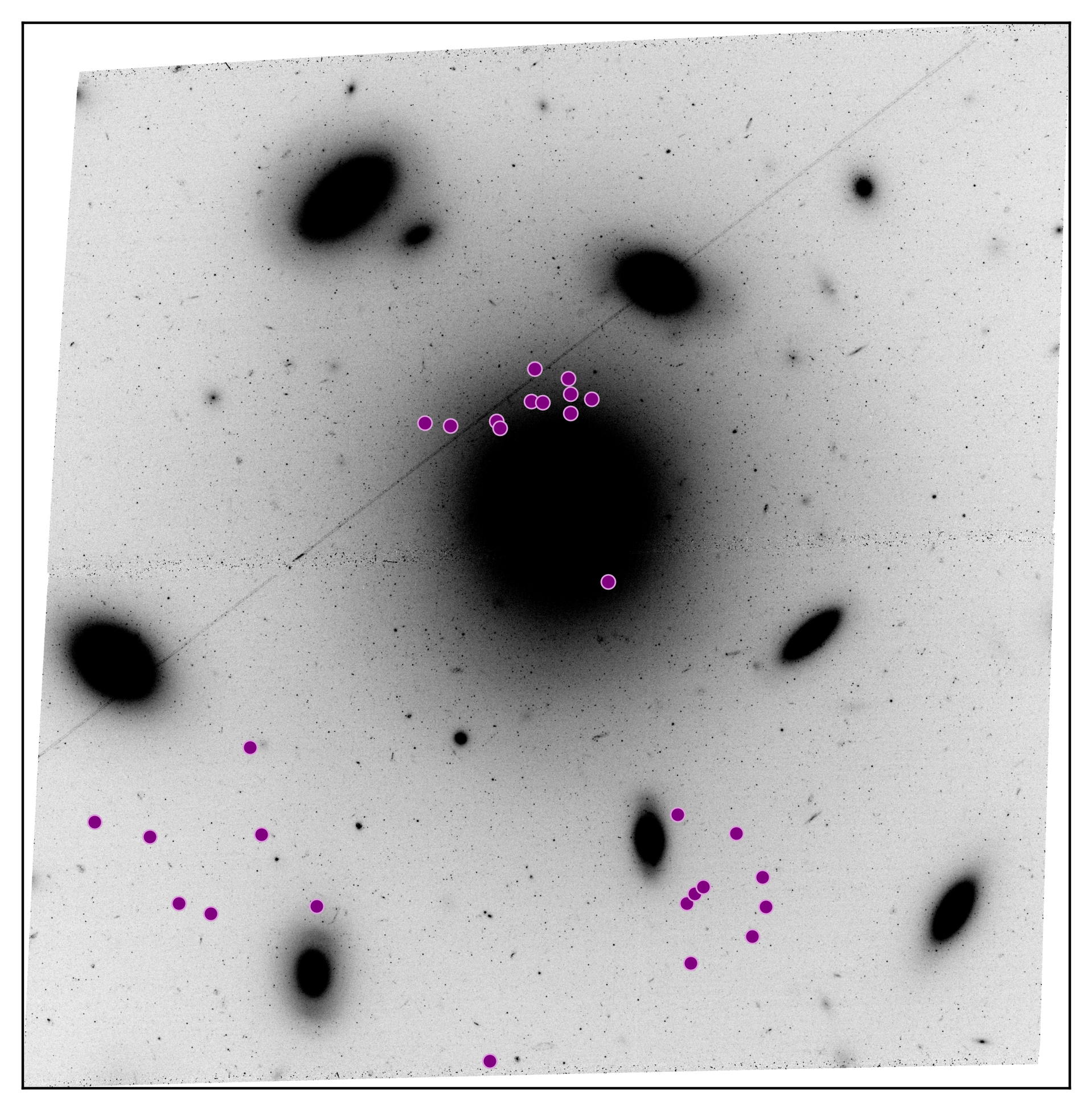}
    \caption{Locations of the 29 GCs in our sample, shown in purple over the F475W reference image of NGC 4874 (large central elliptical galaxy) used in photometry.  WFC3 images cover a smaller field of view than ACS images; selected GCs are therefore confined to the area covered by both instruments.}
    \label{fig:whereGCs}
\end{figure}

\begin{figure}
    \centering
    \includegraphics[width=0.47\textwidth]{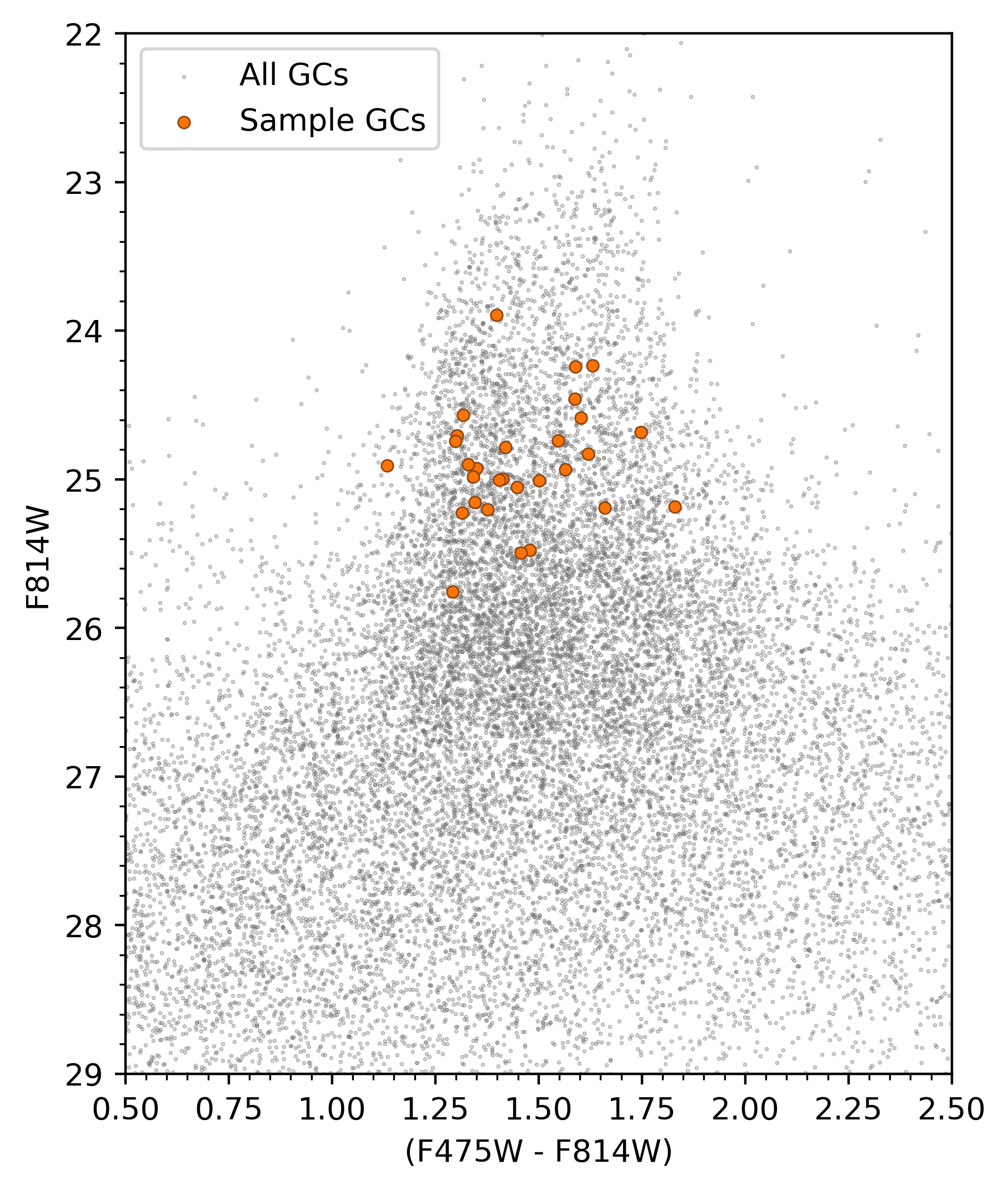}
    \caption{F814W magnitude (F475W - F814W) color for all detected GCs in pale orange, with our sample of 29 bright GCs shown in large, darker orange circles.}
    \label{fig:whereGCscmd}
\end{figure}

NGC 4874 has an extremely rich GCS \citep[see e.g.][]{peng2011richGCS,harris2023colours} and has been imaged in several programs with HST.  In this work we make use of archival observations from GO Program 11711 and new observations from GO Program 17145 for a total of ten available HST filters, from both the ACS (Advanced Camera for Surveys) and WFC3 cameras.  Information about each filter and its corresponding observations can be found in Table \ref{tab:observations}.  This table includes exposure times, extinctions, zeropoints, and a comparison between pivot wavelengths (taken from the \textit{HST} Instrument Handbooks) and effective wavelengths calculated from filter transmission curves following \citet{tokunaga2005effwave} for each filter; the differences between pivot and effective wavelengths are minimal, of order 1 - 10 nm, and have a negligible effect on our fitting procedure.  Because NGC 4874 is distant enough that its GCs appear as point sources in HST images, we were able to use DOLPHOT \citep{dolphin2000dolphot} and its library of HST-specific point spread functions to perform photometry on our images.  From the DOLPHOT output, we selected a sample of 29 GCs using cleaning criteria based on those from \cite{hartman2023enviro}:

\begin{itemize}
    \item DOLPHOT object type 1
    \item Magnitude $< 90$ in all filters
    \item Signal-to-noise ratio $\geq 5$ in all filters
    \item Chi $\leq 1.6$ in all filters
    \item $|\textrm{Sharpness}| < 2.1(0.372 - 0.0390(\textrm{mag}) + 0.001063(\textrm{mag})^2$ for F435W, F438W, F475W, and F475X
    \item $|\textrm{Sharpness}| < 2.4(0.365 - 0.0390(\textrm{mag}) + 0.001063(\textrm{mag})^2$ for F555W, F606W, F814W, F850LP, F110W, F160W
\end{itemize}
These criteria ensured that background galaxies, cosmic ray strikes, bad pixels, and other contaminants, as well as clusters too faint to be reliably measured, were removed from the dataset.  The 29 GCs with high quality measurements in all 10 filters are shown overlaid on NGC 4874 in Figure \ref{fig:whereGCs}, and in a color-magnitude diagram in Figure \ref{fig:whereGCscmd}.

Finally, after compiling our sample, we subtracted foreground reddening \citep{schlafly2011reddening} and converted DOLPHOT's VEGAMAG outputs to flux-based AB magnitudes using the values in Table \ref{tab:observations}.  At the small redshift of the Coma cluster, cosmological K-corrections  to the magnitudes \citep{reinacampos2024rescuer} are $\sim 0.02$ mag or less, within photometric magnitude error, and were not applied.

\begin{table*}
    \begin{threeparttable}
	   \caption{Observations}
	   \label{tab:observations}
      \scriptsize
	   \begin{tabular}{lccccccccc}
            \hline
            Filter & Instrument & Program & Date & \shortstack{Exposure \\ time (s)} & \shortstack{Pivot \\ wavelength (Å)} & \shortstack{Effective \\ wavelength (Å)} & Extinction & \shortstack{VEGAMAG \\ zeropoint} & \shortstack{ABmag \\ zeropoint} \\
            \hline
            F435W & ACS & 17145 & 2024-03-15 & 4563 & 4329.8667 & 4387.0783 & 0.033 & 25.762 & 25.647 \\
            F438W & WFC3 & 17145 & 2024-01-30 & 2505 & 4325.688 & 4348.7982 & 0.033 & 25.002 & 24.837 \\
            F475W & ACS & 11711 & 2010-02-26 & 2394 & 4746.957 & 4792.0191 & 0.030 & 26.156 & 26.049 \\
            F475X & WFC3 & 17145 & 2024-03-21 & 2496 & 4939.0691 & 5040.8426 & 0.028 & 26.213 & 26.156 \\
            F555W & ACS & 17145 & 2023-12-16 & 4563 & 5361.0298 & 5370.8398 & 0.025 & 25.710 & 25.699 \\
            F606W & ACS & 17145 & 2024-01-07 & 2235 & 5921.8931 & 5933.2926 & 0.023 & 26.397 & 26.484 \\
            F814W & ACS & 11711 & 2010-02-06 & 10425 & 8045.5298 & 7993.8419 & 0.014 & 25.510 & 25.941 \\
            F850LP & ACS & 17145 & 2024-03-18 & 4563 & 9031.4619 & 8993.2430 & 0.011 & 24.318 & 24.846 \\
            F110W & WFC3 & 11711 & 2010-03-01 & 3197 & 11534.459 & 11342.175 & 0.008 & 26.042 & 26.829 \\
            F160W & WFC3 & 11711 & 2010-03-01 & 3597 & 15369.176 & 15324.969 & 0.005 & 24.662 & 25.936 \\
            \hline
	    \end{tabular}
        \begin{tablenotes}
            \item Columns: (1) Filter name; (2) \textit{HST} camera; (3) HST GO program number associated with each filter; (4) Calendar date of observations in each filter; (5) Total exposure time; (6) Central wavelength of each filter (citations)---in plots of a quantity vs. wavelength, values derived from each filter will appear at the pivot wavelength; (7) Foreground extinction from \cite{schlafly2011reddening}; (8) Zeropoint in the VEGAMAG system used by DOLPHOT; (9) Zeropoint in the ABmag system used in this work.  \textbf{Pivot wavelengths are from the \href{https://hst-docs.stsci.edu/acsihb/chapter-5-imaging/5-1-imaging-overview}{ACS} and \href{https://hst-docs.stsci.edu/wfc3ihb/chapter-6-uvis-imaging-with-wfc3/6-5-uvis-spectral-elements}{WFC3 Instrument Handbooks}; effective wavelengths were calculated using the filter transmission curves available in the Instrument Handbooks and a BaSTI [Fe/H] = -0.96, 12.5 Gyr isochrone, representing a GC of middling metallicity and age.}  WFC3 zeropoints are from \cite{calamida2022zerosWFC3}; ACS zeropoints were calculated using the ACS Team's \href{https://acszeropoints.stsci.edu/}{Zeropoint Calculator webtool}.
        \end{tablenotes}
    \end{threeparttable}
\end{table*}

\subsection{Model spectra} \label{subsec:EMILES}

We used a subset of the E-MILES single stellar population (SSP) library \citep{vazdekis2016EMILES}, comprising the range of metallicities and ages typical for GCs.  E-MILES SEDs are derived from Padova \citep{girardi2000padova} and BaSTI \citep{pietrinferni2004basti} isochrones, and they have a range of discrete possible values and options for metallicity, age, and IMF type and slope.  The parameters for our subset library of SEDs are listed in Table \ref{tab:SEDparams}.  We included all metallicities and ages that are plausible for GCs and three IMF options.

\begin{table*}
	\begin{threeparttable}
	    \centering
	    \caption{Model spectra specifications}
	    \label{tab:SEDparams}
	    \begin{tabular}{ll|l}
		    \hline
		    Isochrone & Parameter & Input(s) \\
            \hline
		    BaSTI & Type of IMF & \textbf{Kroupa universal}, Chabrier, Salpeter \\
		     & IMF slope & \textbf{1.30} \\
             & Metallicity [M/H] & \textbf{-2.27, -1.79, -1.49, -1.26, -0.96, -0.66, -0.35, -0.25, +0.06} \\
             & Age (Gyr) & 9.0, 9.5, 10.0, 10.5, 11.0, 11.5, 12.0, \textbf{12.5}, 13.0, 13.5 \\
		    \hline
            Padova & Type of IMF & \textbf{Kroupa universal}, Chabrier, Salpeter \\
		     & IMF slope & \textbf{1.30} \\
             & Metallicity [M/H] & \textbf{-2.32, -1.71, -1.31, -0.71, -0.40, 0.00} \\
             & Age (Gyr) & 8.9125, 10.0000, 11.2202, \textbf{12.5893}, 14.1254 \\
            \hline
	    \end{tabular}
        \begin{tablenotes}
            \item Customizable properties of the E-MILES model spectra.  Parameters used in the nominal fitting routine are shown in bold.
        \end{tablenotes}
	\end{threeparttable}
\end{table*}

\section{SED fitting process} \label{sec:fitting}

\subsection{Flux to AB magnitude} \label{subsec:flux2AB}

The E-MILES models give flux $F_{\lambda}$ versus wavelength $\lambda$ in units of Solar luminosities per Ångstrom per Solar mass.  While model SEDs may differ systematically from real stellar spectra, they have no measurement uncertainty---therefore, in order to avoid nonlinear effects that would stem from propagating our photometry measurement errors from magnitude to flux, we instead transformed the SEDs from units of flux to AB magnitude and fitted the SEDs in AB space.  In order to convert the model spectra to AB magnitude, we need to account for the distance to our GCs to be fitted and the total cluster mass:

\begin{equation} \label{eq:Lsun2flamb}
    f_{\lambda} = \frac{M_{GC}L_{\odot}}{4 \pi D^2}F_{\lambda}
\end{equation}
where $f_{\lambda}$ is the measured flux for the cluster of mass $M_{GC}$ at distance $D$.  For Coma we adopt a luminosity distance $D = 109$ Mpc (from NED).

After converting the E-MILES fluxes to cgs units, we transform $f_{\lambda}$ to $f_{\nu}$, in frequency units:
\begin{equation} \label{eq:flamb2fnu}
    f_{\nu} = \frac{\lambda^2}{c}f_{\lambda} = 3.34 \times 10^4 \lambda^2 f_{\lambda} \textrm{ Jy}
\end{equation}
where now $\lambda$ is in Ångstroms and $f_{\nu}$ in Janskys.

With flux in terms of frequency, we find the equivalent magnitude in the frequency-based AB magnitude system:
\begin{equation} \label{eq:fnu2AB}
    m_{AB} = 8.90 - 2.5\textrm{log}(f_{\nu})
\end{equation}

Combining Equations \ref{eq:Lsun2flamb}, \ref{eq:flamb2fnu}, and \ref{eq:fnu2AB} makes the dependence of AB magnitude on GC mass explicit:
\begin{equation} \label{eq:GCmassdep}
    m_{AB} = const - 2.5\textrm{log}(\frac{\lambda^2}{4 \pi D^2}F_{\lambda}) - 2.5\textrm{log}(M_{GC})
\end{equation}

Finally, after finding the constant appropriate for the convenient units of Mpc for distance, Å for wavelength, $L_{\odot}$/Å/$M_{\odot}$ for E-MILES flux, and $M_{\odot}$ for GC mass, we can go directly from E-MILES flux to AB magnitude, with each input quantity from Equation \ref{eq:GCmassdep} isolated to illustrate its affect on the resulting magnitude:
\begin{multline} \label{eq:Lsun2AB}
    m_{AB} = 38.226 + 5\textrm{log}(D) - 5\textrm{log}(\lambda) \\ - 2.5\textrm{log}(F_{\lambda}) - 2.5\textrm{log}(M_{GC})
\end{multline}

\subsection{Fitting routine} \label{subsec:fitting}

Using Equation \ref{eq:Lsun2AB}, we implemented a chi-squared minimization routine to fit E-MILES SEDs to the observed GC AB magnitudes, with photometric magnitude error weighting.  Previous discussions of SED fitting to integrated GC photometry \citep[e.g.][]{faisst2022tinythings} have tended to concentrate on cluster metallicity and age, though as can be seen from the outline above, cluster mass and thus mass-to-light ratio is an automatic and equally simple result of the fitting process (see below), particularly when the clusters are old enough that age has a negligible effect on SED shape (more discussion below in Section \ref{subsec:age_fits}).

The routine selects the best fitting SED from a library spanning the parameters outlined in bold in Table \ref{tab:SEDparams} and ranging in cluster mass from $10^5$ to $10^7 \textrm{M}_{\odot}$ (i.e. plausible masses for GCs).  We also created alternate versions of the routine, assuming different GC ages and different IMF configurations, to test the assumptions made in our nominal routine.  When finding the best fit, we do not attempt to interpolate in metallicity or age between the individual values listed in Table \ref{tab:SEDparams}.

\subsection{Age and metallicity considerations} \label{subsec:agemet}

\begin{figure}
    \centering
    \includegraphics[width=0.47\textwidth]{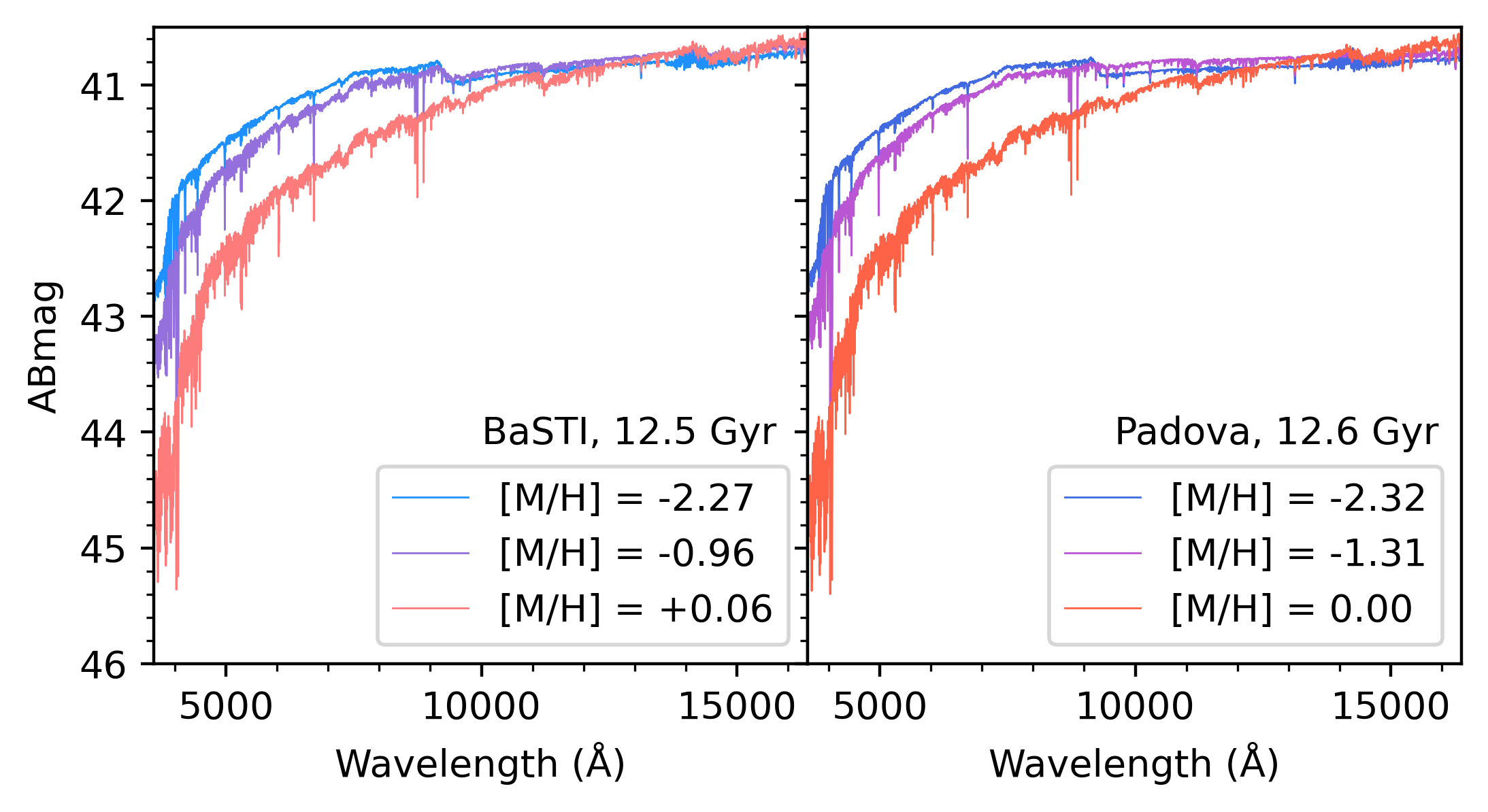}
    \caption{A comparison of model spectra of different metallicities at fixed age for BaSTI (left) and Padova (right), from the original 1-M$\odot$ E-MILES outputs.  Changing the metallicity input affects the shape of the spectrum, with lower metallicity spectra being flatter (i.e. less difference between magnitudes at different wavelengths) and higher metallicity spectra being more sloped (i.e. greater differences between the fainter short wavelengths and brighter long wavelengths).  Metal-poor spectra are shown in blue, intermediate metallicity in purple, and metal-rich in red.}  
    \label{fig:metcomp}
\end{figure}

\begin{figure}
    \centering
    \includegraphics[width=0.47\textwidth]{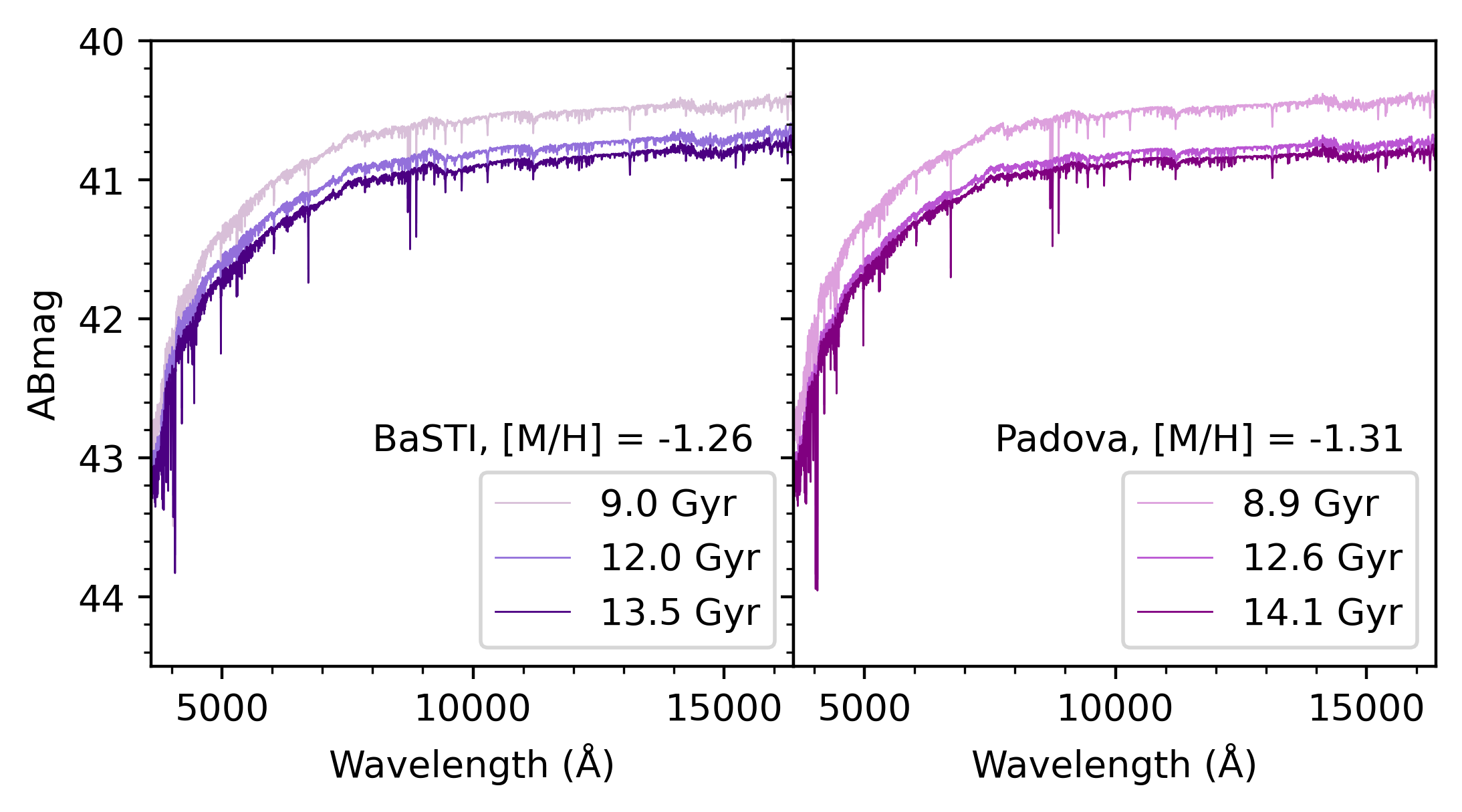}
    \caption{A comparison of model spectra of different ages at fixed metallicity for BaSTI (left) and Padova (right), from the original 1-M$\odot$ E-MILES outputs.  While the brightness of the spectra changes with age, their shapes are largely unaffected.  "Young" (for GCs) spectra are shown in light purple, middling age in medium purple, and old in dark purple.}
    \label{fig:agecomp}
\end{figure}

Our nominal fitting routine fixes GC age while fitting for metallicity and mass.  This configuration was motivated by the differing effects of metallicity and age on GC photometry.  While metallicity affects the color (i.e. the relative photometric magnitudes at different wavelengths) of a GC \citep[see e.g.][]{peng2006nonlinear,sinnott2010colour,usher2012colour,harris2017BCGs,fahrion2020spectroscopy}, age has a much smaller effect on GC observables on objects this old.  As a GC ages beyond about 8 Gyr, the position of the red-giant branch in the cluster CMD  (color-magnitude diagram) that produces most of the integrated light changes extremely slowly; we therefore expect age to have a negligible effect on GC magnitudes compared to metallicity (see \cite{hartman2024models} for a more detailed discussion of these effects in color-color space).  Figures \ref{fig:metcomp} and \ref{fig:agecomp} more clearly illustrate the effects of metallicity and age, respectively, on GC SEDs: a change in metallicity affects the slope of the SED, particularly at shorter, bluer wavelengths, while a change in age affects only the overall brightness of the SED (by less than half a magnitude over our entire plausible GC age range).

\subsection{Other derived quantities} \label{subsec:lums}

One potential use for fitted GC masses aside from a plausible mass reality check is in inferring GC mass-to-light (M/L) ratios.  To that end, we derived four wavelength-specific luminosities for each sample GC, using magnitudes from red and near-infrared filters in the flat (and therefore better-constrained) portion of the GC SED:

\begin{equation} \label{eq:luminosity}
    -2.5\textrm{log}(L_{\lambda}) = \textrm{m}_{AB,\lambda} - 5\textrm{log}(D) - \textrm{M}_{\odot,\lambda}
\end{equation}
where m$_{AB,\lambda}$ is the measured AB magnitude in a particular filter, M$_{\odot,\lambda}$ is the absolute AB magnitude of the Sun in that filter \citep[see][Table 3 for the values used in this work]{willmer2018solarmags}, and $D$ is the luminosity distance to the cluster in units of 10 pc (if using a luminosity distance in e.g. Mpc, multiply by $10^5$).

\begin{table}
    \begin{threeparttable}
       \centering
	   \caption{Solar magnitudes}
	   \label{tab:solarmags}
	   \begin{tabular}{ccc}
            \hline
            Filter & AB Magnitude & VEGAMAG \\
            \hline
            F814W & 4.52 & 4.12 \\
            F850LP & 4.50 & 4.01 \\
            F110W & 4.52 & 3.75 \\
            F160W & 4.60 & 3.37 \\
            \hline
	    \end{tabular}
        \begin{tablenotes}
            \item Solar absolute magnitudes used in Equation \ref{eq:luminosity}.  Columns: (1) Filter name; (2) Absolute Solar magnitude in the AB system; (3) Absolute Solar magnitude in the VEGAMAG system.  Magnitudes are from \cite{willmer2018solarmags}.  The magnitudes for F110W have been interpolated.
        \end{tablenotes}
    \end{threeparttable}
\end{table}

\section{Results} \label{sec:results}

\subsection{Nominal fits and comparison to local GCs} \label{subsec:all_fits}

\begin{figure*}
    \centering
    \includegraphics[width=0.97\textwidth]{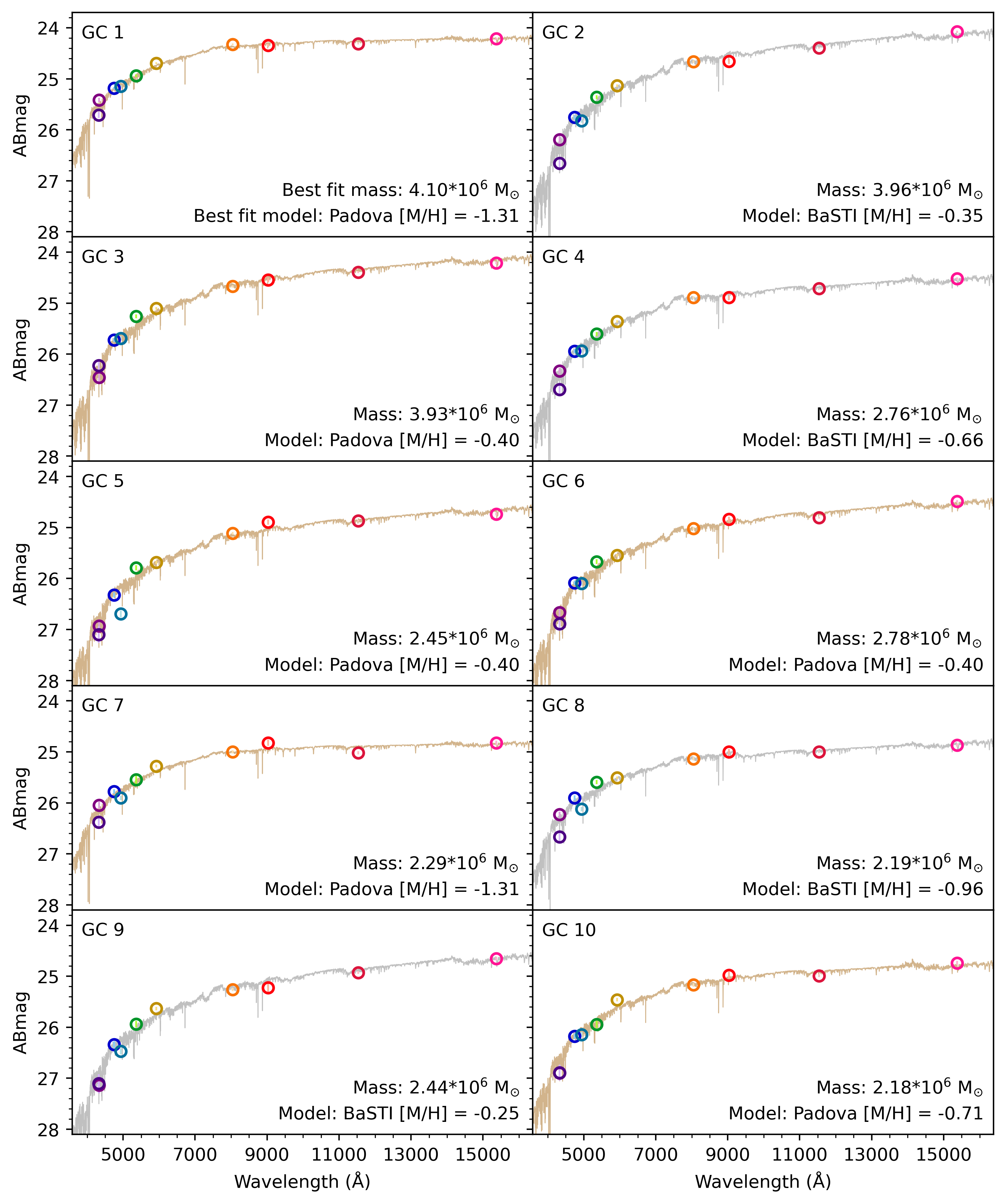}
    \caption{Magnitudes and fitted SEDs for sample GCs.  Fitted masses, models, and metallicites are listed in the bottom right of each panel.  BaSTI SEDs are shown in grey and Padova SEDs in tan.  Magnitudes are color-coded as follows: purple for F435W, indigo for F438W, blue for F475W, teal for F475X, green for F555W, yellow for F606W, orange for F814W, bright red for F850LP, dark red for F110W, and pink for F160W.}
    \label{fig:GCfits1}
\end{figure*}

\begin{figure*}
    \centering
    \includegraphics[width=0.97\textwidth]{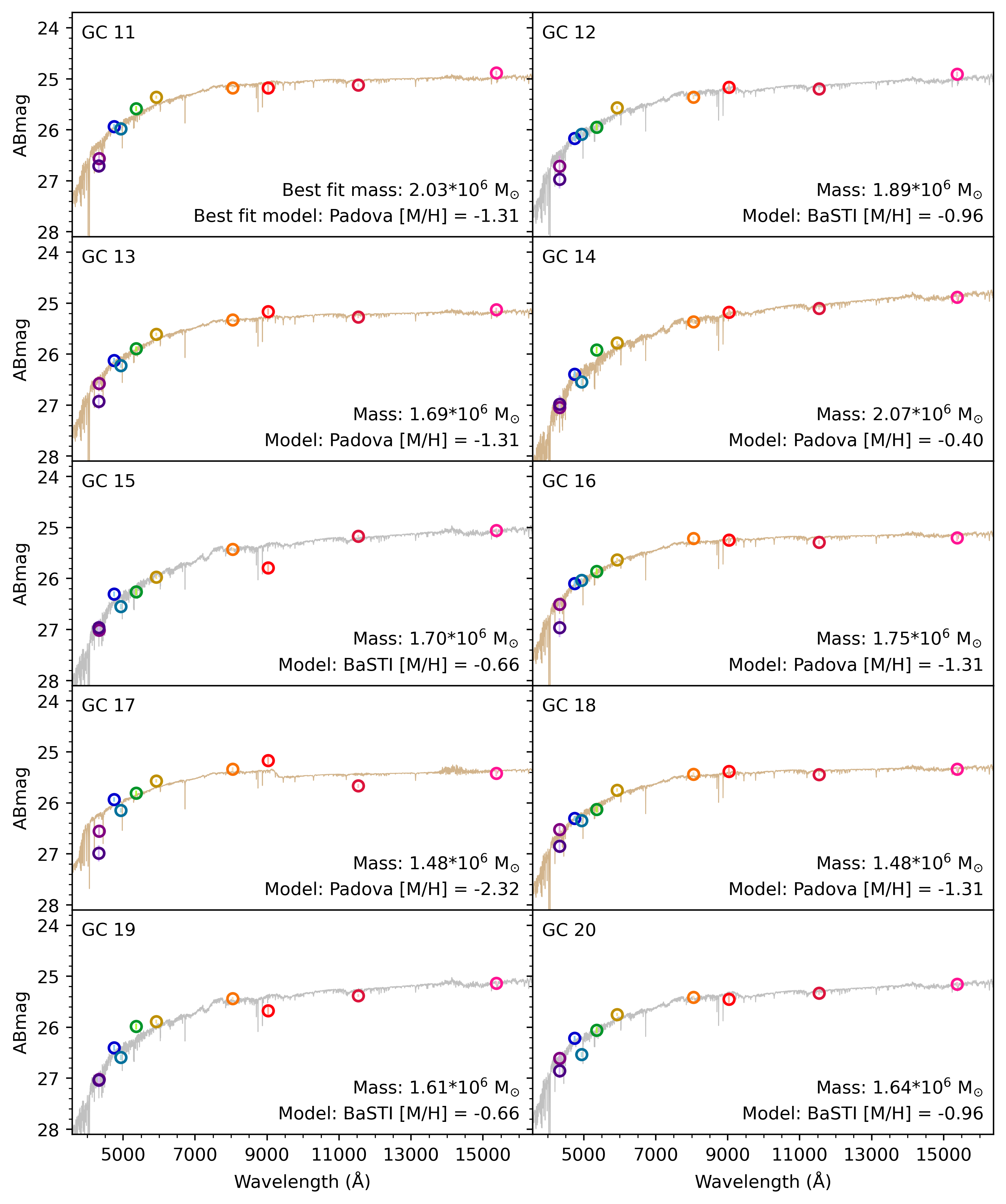}
    \caption{Same as Figure \ref{fig:GCfits1}.}
    \label{fig:GCfits2}
\end{figure*}

\begin{figure*}
    \centering
    \includegraphics[width=0.97\textwidth]{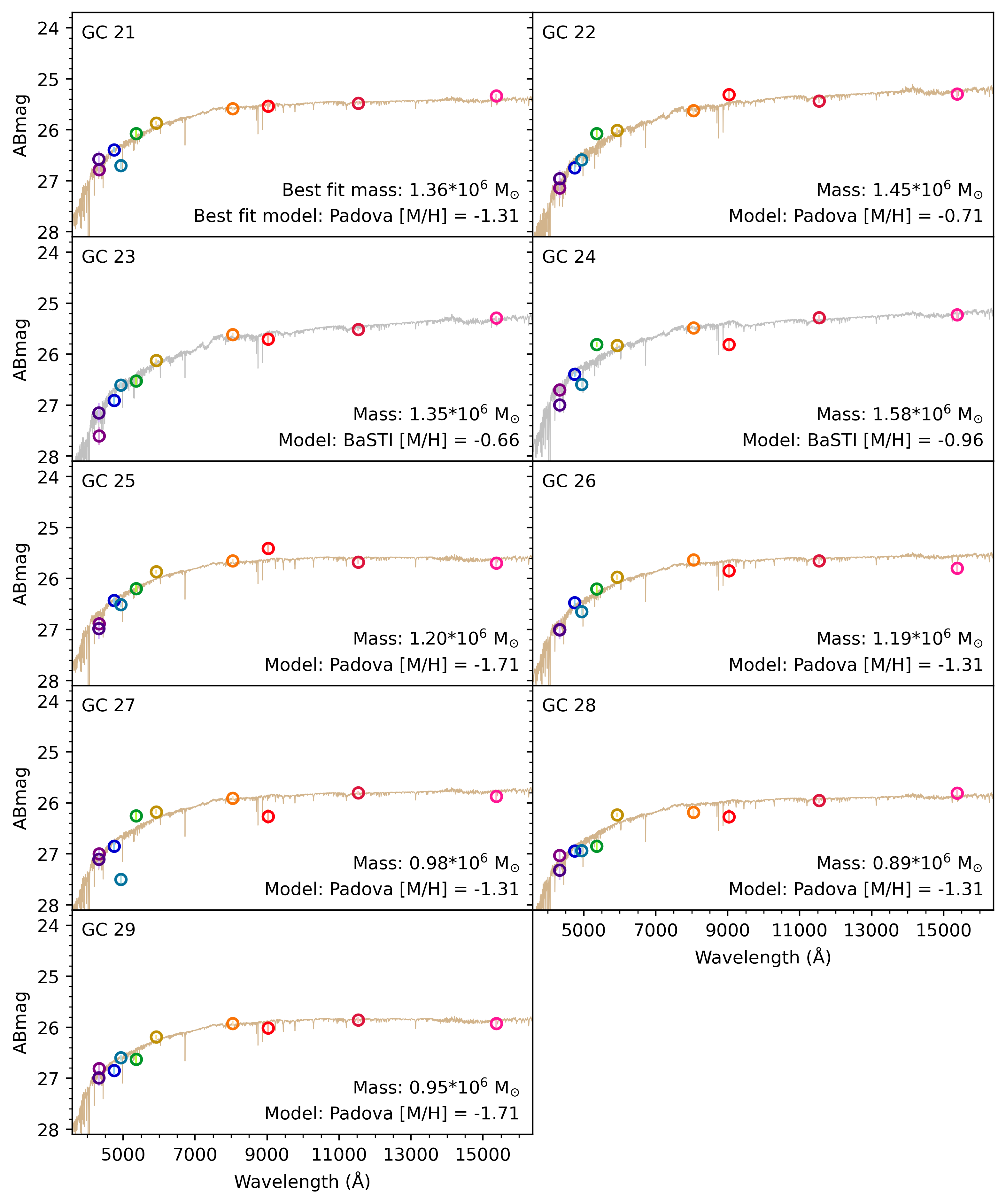}
    \caption{Same as Figure \ref{fig:GCfits1}.}
    \label{fig:GCfits3}
\end{figure*}

Figures \ref{fig:GCfits1}, \ref{fig:GCfits2}, and \ref{fig:GCfits3} show SEDs from our nominal fitting routine overplotted with photometry from all 29 GCs in our test sample.  HST filters are color-coded, and fitted masses and metallicities are listed for each GC.

\begin{figure}
    \centering
    \includegraphics[width=0.47\textwidth]{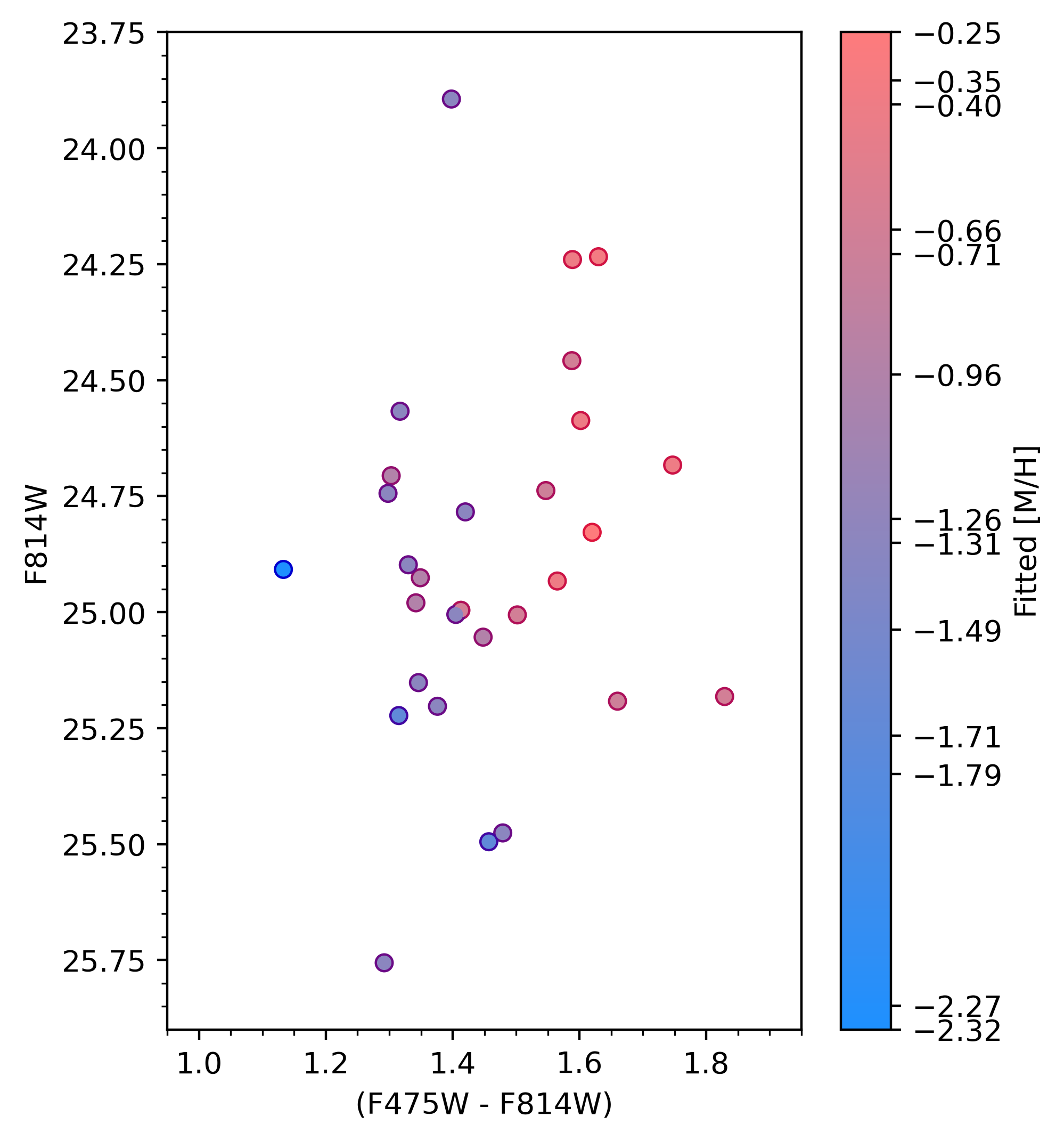}
    \caption{F814W magnitude vs. (F475W - F814W) color for the sample GCs.  Points are colored by fitted metallicity.  As expected, blue GCs tend to be metal-poor and red GCs tend to be metal-rich.}
    \label{fig:metCMD}
\end{figure}

All fitted metallicities are sub-Solar, in line with previous observations of GCs in many other galaxies \citep[e.g.][]{peng2006nonlinear,woodley2010agemet,pastorello2015met,fahrion2020spectroscopy,harris2023colours}, and only one of the sample GCs was fitted with the most metal-poor SED at [Fe/H] = -2.32, comparable to the most metal-poor Galactic and local group GCs \citep[see e.g.][]{harris1996catalogue,harris2010catalogue,larsen2022metallicities}.  In a CMD (Figure \ref{fig:metCMD}), fitted metallicities appear as expected, with blue GCs predominantly metal-poor and red GCs predominantly metal-rich \citep[see][among others]{sinnott2010colour,usher2012colour,fahrion2020spectroscopy,harris2023colours,hartman2024models}.

Fitted masses run from $8.7 \times 10^5 \textrm{M}_{\odot}$ to $4.1 \times 10^6 \textrm{M}_{\odot}$.  These masses are on the high end of the normal GC mass range \citep{harris2014gclum}, but fall within the range of dynamical masses measured in \cite{strader2011m31}.

\begin{figure*}
    \centering
    \includegraphics[width=0.97\textwidth]{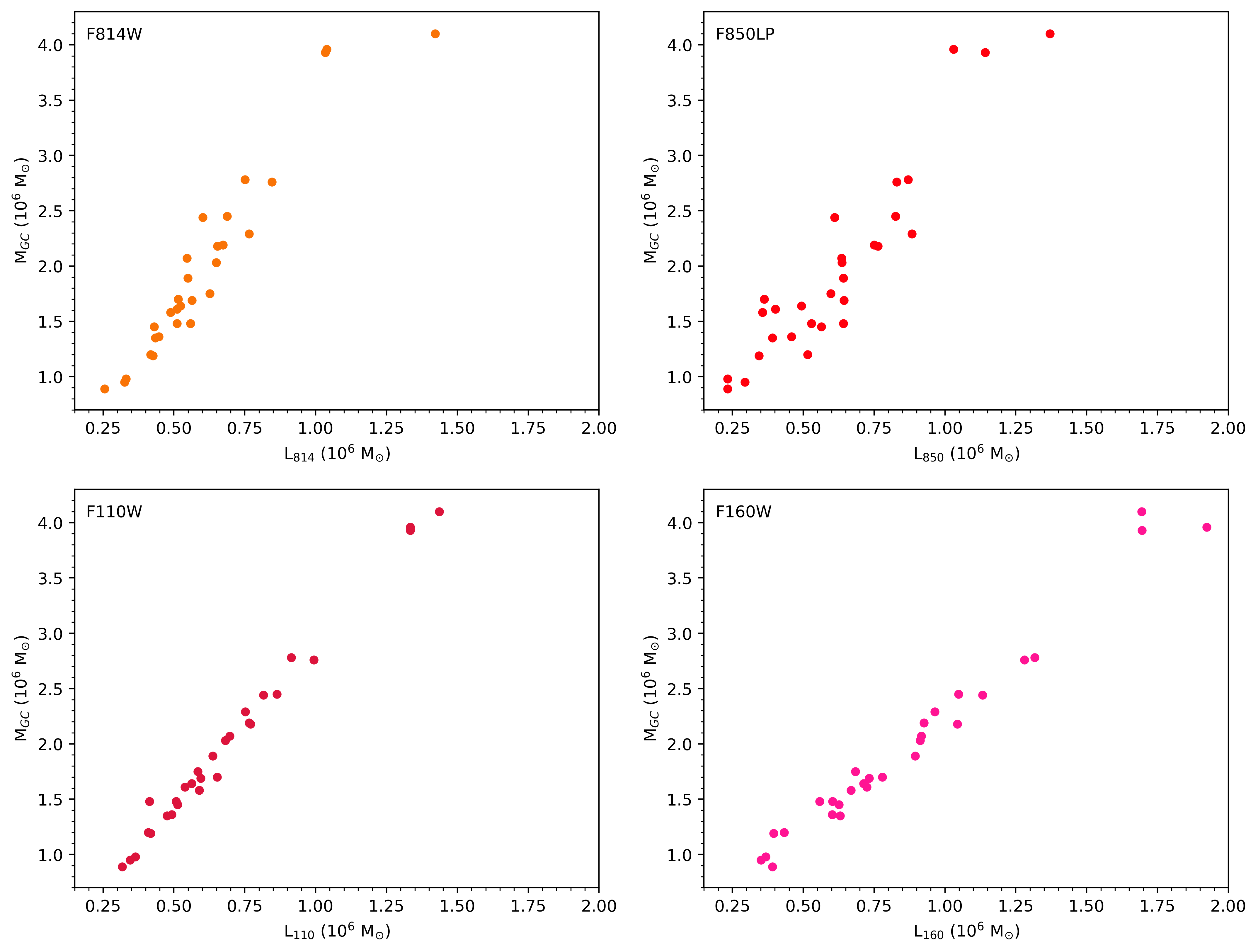}
    \caption{Fitted mass vs. luminosity for all 29 sample GCs, with luminosity derived from F814W (top left, orange), F850LP (top right, bright red), F110W (bottom left, dark red), and F160W (bottom right, pink) magnitudes according to Equation \ref{eq:luminosity} and Table \ref{tab:solarmags}.}
    \label{fig:MvsL}
\end{figure*}

\begin{figure*}
    \centering
    \includegraphics[width=0.95\textwidth]{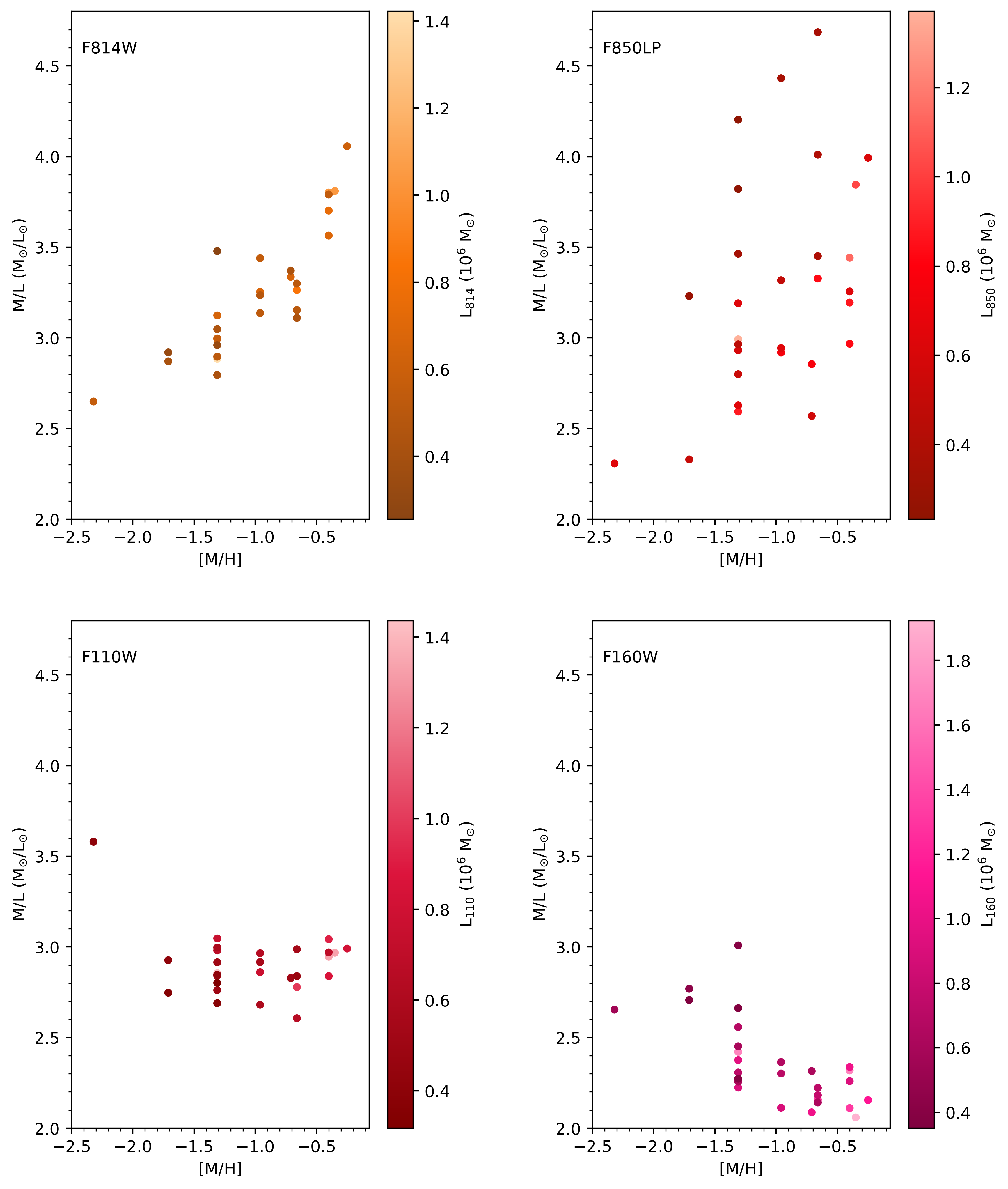}
    \caption{Derived M/L ratio vs. fitted metallicity for all 29 sample GCs, with luminosity derived from F814W (top left, orange), F850LP (top right, bright red), F110W (bottom left, dark red), and F160W (bottom right, pink) magnitudes according to Equation \ref{eq:luminosity} and Table \ref{tab:solarmags}.  Lighter points denote brighter GCs and darker points fainter GCs.  Because F814W sits near where the spectra begin to turn over at $z = 0.024$, there is a correlation between M/L ratio (through magnitude) and metallicity (i.e. spectrum shape) for M/L ratios derived from this filter.  This correlation disappears for redder filters, which fall within the flat portion of the spectra.}
    \label{fig:MLvsFeH}
\end{figure*}

Lastly, using our fitted GC masses and derived luminosities, we found the mass-to-light (M/L) ratio for each GC in our sample.  Figure \ref{fig:MvsL} shows GC mass vs. luminosity, for the four reddest magnitudes (F814W, F850LP, F110W, and F160W covering the flat part of the SED), and listed in Table \ref{tab:ML_fits}.  Our derived M/L ratios, mostly in the range $\simeq 2 - 4 M_{\odot}/L_{\odot}$, tend to be higher than those of Galactic GCs at lower mass ($\lesssim 10^5 M_{\odot}$)\citep{baumgardt2020masstolight,harris2017masstolight}, but within the range for GCs in other galaxies \citep[e.g.][]{strader2011m31,dumont2022masstolight}.  There is also some evidence that M/L increases with GC mass above $10^5 M_{\odot}$ \citep[see][for discussion]{strader2011m31,harris2017masstolight}. M/L ratios based on longer wavelengths may be more reliable than their shorter wavelength counterparts---shorter wavelength filters fall within the downturned, highly metallicity-dependent portion of the GC spectrum, while longer wavelength filters fall within the flat, less metallicity-dependent portion.  Figure \ref{fig:MLvsFeH} compares M/L ratios based on photometry in the F814W filter, which includes the beginning of the downturn, to those derived from F850LP, F110W, and F160W; the F814W M/L ratios show a correlation with metallicity that is not present in the other three filters.

\begin{table*}
    \begin{threeparttable}
       \centering
	   \caption{Derived masses and M/L and positions}
	   \label{tab:ML_fits}
	   \begin{tabular}{l|ccccccc}
            \hline
             GC & R.A. (hms)  & Dec. (dms) & Fitted mass ($10^6 M_{\odot}$) & $M/L_{814}$ & $M/L_{850}$ & $M/L_{110}$ & $M/L_{160}$ \\
            \hline
            GC1  & 12:59:36.2301 & +27:59:02.775 &  4.10 & 2.88 & 2.99 & 2.85 & 2.42 \\
            GC2  & 12:59:33.5895 & +27:57:38.941 &  3.96 & 3.81 & 3.84 & 2.97 & 2.06 \\
            GC3  & 12:59:33.8253 & +27:57:18.670 &  3.93 & 3.80 & 3.44 & 2.95 & 2.32 \\
            GC4  & 12:59:42.0863 & +27:58:16.463 &  2.76 & 3.26 & 3.33 & 2.78 & 2.16 \\
            GC5  & 12:59:36.5968 & +27:59:21.063 &  2.45 & 3.56 & 2.97 & 2.84 & 2.34 \\
            GC6  & 12:59:41.3812 & +27:58:08.953 &  2.78 & 3.70 & 3.19 & 3.04 & 2.11 \\
            GC7  & 12:59:34.7399 & +27:57:14.311 &  2.29 & 2.99 & 2.59 & 3.05 & 2.38 \\
            GC8  & 12:59:41.1024 & +27:57:52.084 &  2.19 & 3.25 & 2.92 & 2.86 & 2.36 \\
            GC9  & 12:59:34.2778 & +27:57:14.991 &  2.44 & 4.06 & 3.99 & 2.99 & 2.15 \\
            GC10 & 12:59:34.1517 & +27:57:23.576 &  2.18 & 3.34 & 2.85 & 2.83 & 2.09 \\
            GC11 & 12:59:34.4751 & +27:57:16.759 &  2.03 & 3.12 & 3.19 & 2.98 & 2.22 \\
            GC12 & 12:59:34.0148 & +27:57:31.532 &  1.89 & 3.44 & 2.94 & 2.96 & 2.11 \\
            GC13 & 12:59:42.2318 & +27:57:57.629 &  1.69 & 3.00 & 2.63 & 2.84 & 2.31 \\
            GC14 & 12:59:34.1218 & +27:57:31.884 &  2.07 & 3.79 & 3.26 & 2.97 & 2.26 \\
            GC15 & 12:59:41.8632 & +27:57:54.113 &  1.70 & 3.30 & 4.69 & 2.61 & 2.18 \\
            GC16 & 12:59:36.9575 & +27:57:36.787 &  1.75 & 2.79 & 2.93 & 3.00 & 2.56 \\
            GC17 & 12:59:33.2979 & +27:57:42.377 &  1.48 & 2.65 & 2.31 & 3.58 & 2.65 \\
            GC18 & 12:59:41.3510 & +27:58:06.487 &  1.48 & 2.90 & 2.80 & 2.91 & 2.45 \\
            GC19 & 12:59:35.1330 & +27:58:52.559 &  1.61 & 3.15 & 4.01 & 2.99 & 2.22 \\
            GC20 & 12:59:36.1489 & +27:59:24.364 &  1.64 & 3.14 & 3.32 & 2.92 & 2.30 \\
            GC21 & 12:59:37.6059 & +27:59:04.287 &  1.46 & 3.05 & 2.96 & 2.76 & 2.26 \\
            GC22 & 12:59:34.2795 & +27:57:22.046 &  1.45 & 3.37 & 2.57 & 2.83 & 2.31 \\
            GC23 & 12:59:41.3612 & +27:58:04.193 &  1.35 & 3.11 & 3.45 & 2.84 & 2.14 \\
            GC24 & 12:59:34.3754 & +27:59:25.773 &  1.58 & 3.23 & 4.43 & 2.68 & 2.36 \\
            GC25 & 12:59:40.2868 & +27:57:58.162 &  1.20 & 2.87 & 2.33 & 2.93 & 2.77 \\
            GC26 & 12:59:34.6886 & +27:57:19.418 &  1.19 & 2.79 & 3.46 & 2.85 & 3.01 \\
            GC27 & 12:59:35.1064 & +27:59:19.557 &  0.98 & 2.96 & 4.20 & 2.69 & 2.66 \\
            GC28 & 12:59:41.1140 & +27:58:59.708 &  0.89 & 3.48 & 3.82 & 2.80 & 2.27 \\
            GC29 & 12:59:42.4168 & +27:58:03.741 &  0.95 & 2.92 & 3.23 & 2.75 & 2.71 \\
            \hline
	    \end{tabular}
        \begin{tablenotes}
            \item Columns: (1) Sample GC identifier; (2-3) Equatorial position; (4) Fitted mass; (5) M/L from F814W, (6) F850LP, (7) F110W, and (8) F160W.  All M/L ratios in units of $M_{\odot}/L_{\odot}$.
        \end{tablenotes}
    \end{threeparttable}
\end{table*}

\subsection{The effect of age and IMF} \label{subsec:age_fits}

In addition to the nominal fitting routine, which assumes a GC age of $\sim$ 12.5 Gyr and adopts a Kroupa universal IMF \citep{kroupa1993imf}, we tested a set of routines that assumed other age brackets, and two routines that assumed alternate IMF shapes.

As expected, changing the assumed age of the GCs had little affect on the fitted metallicity---in roughly 75\% of cases, fitting in a different age bracket produced the same metallicity result as the nominal fitting routine.  In cases where the alternate routine returned a different metallicity, the largest discrepancies were two metallicity bins away from the nominal result.  This lack of effect on metallicity results was expected; once a GC ages beyond ~8 Gyr, the position of the red giant branch on a CMD, which affects its colour indices and SED shape, changes so slowly that it is negligible in comparison to the effect of metallicity itself.  Fitted masses varied systematically by age bracket on the order of $10^5 \textrm{M}_{\odot}$, with the youngest bracket returning the lowest masses and the oldest bracket returning the largest masses.  This effect stems from simple stellar evolution---as the model SSPs age, brighter stars disappear from the cluster; older SSPs require more mass to reach the same luminosity as younger ones.

\begin{figure}
    \centering
    \includegraphics[width=0.47\textwidth]{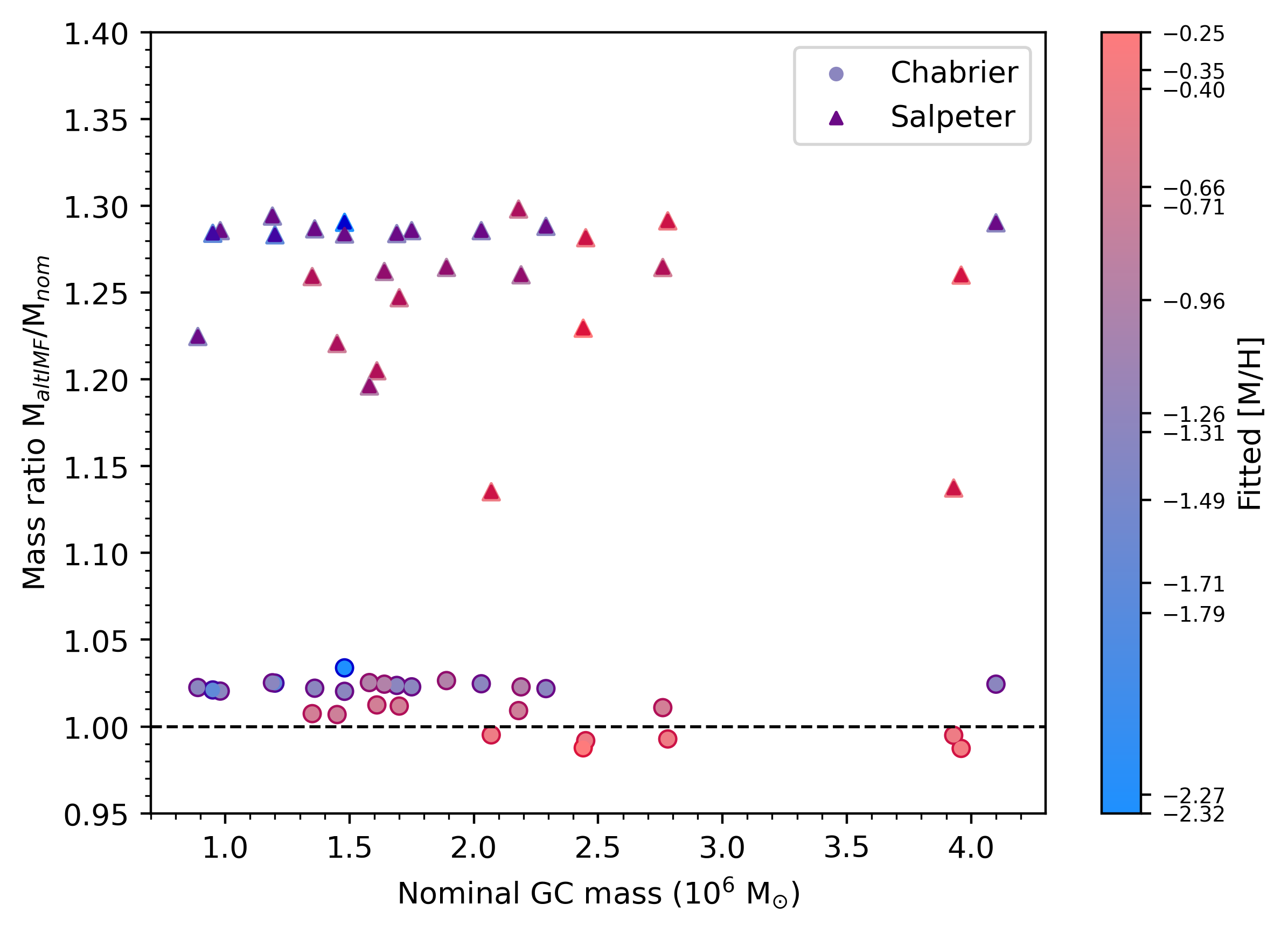}
    \caption{Fitted mass from alternate IMF routines compared to nominal fitted mass as a function of nominal fitted mass.  Results from the Chabrier fitting routine are shown as circles; results from the Salpeter fitting routine are shown as triangles.  Points are colored by nominal fitted metallicity.}
    \label{fig:metIMF}
\end{figure}

Assuming a Chabrier IMF \citep{chabrier2001imf} returned the same metallicity fits and nearly the same masses (within $\pm$ 5\%) as the nominal routine for all 29 GCs.  Assuming a Salpeter IMF \citep{salpeter1955imf} returned the same metallicity as the nominal routine in a slim majority of cases, with discrepancies of a maximum of two metallicity bins.  Fitted masses assuming a Salpeter IMF, which has a higher contribution from low-mass stars than Kroupa or Chabrier, ranged from about 10\% to over 30\% larger than those from the nominal routine.  Figure \ref{fig:metIMF} shows the ratio of fitted masses from the alternate routines compared to the nominal routine as a function of nominal fitted mass.

\subsection{Fitting with fewer filters} \label{subsec:fewer_filts}

Because exceedingly few galaxies have photometry available in as many as  ten bandpasses, we tested our fitting routine with varying combinations of filters removed from the dataset.  How many filters, and which ones, are needed to return reliable results?

We first cumulatively removed filters one at a time in order of wavelength, starting from the bluest available (F435W) in order to constrain where in the GC spectrum we lose leverage on metallicity and on mass, flagging the point at which fitted metallicity changed by more than one model spectrum bin and the point at which fitted mass changed by more than 10\% compared to the nominal ten-filter fitting results.  We then repeated this exercise in the opposite way, starting instead from the reddest available (F160W). Loss of metallicity leverage was most common after the removal of F475X photometry (i.e. fitting with only F555W and redder), and loss of mass leverage was most common after removal of F814W photometry (i.e. fitting with only F606W and bluer).

Given the need for both red and blue filters in order to adequately constrain metallicity and mass, we performed a third exercise: removing filters in a sawtooth or fence-post pattern, keeping only every second filter or every third filter to ensure even coverage of the GC spectrum.  Differences in fitting results between the nominal ten-filter fitting procedure and fitting with sawtooth filter removal are summarized in Table \ref{tab:sawtooth_filts}.

\begin{table*}
    \begin{threeparttable}
       \centering
	   \caption{Filter removal}
	   \label{tab:sawtooth_filts}
	   \begin{tabular}{ccc}
            \hline
            Filters remaining & Average $\Delta M/M$ (\%) & Metallicity change frequency (\%) \\
            \hline
            F435W, F475W, F555W, F814W, F110W & 4.5 & 62 \\
            F438W, F475X, F606W, F850LP, F160W & 3.7 & 41 \\
            F435W, F475X, F814W, F160W & 3.5 & 72 \\
            F438W, F555W, F850LP & 14 & 83 \\
            F475W, F606W, F110W & 4.7 & 79 \\
            \hline
	    \end{tabular}
        \begin{tablenotes}
            \item Differences in fitting results with fewer than ten filters.  Columns: (1) Filters used in fitting; (2) Average fractional change in fitted mass across the sample of 29 GCs; (3) Frequency of change in metallicity model (i.e. how often did the fitting routine return a different metallicity than the nominal ten-filter fit?).
        \end{tablenotes}
    \end{threeparttable}
\end{table*}

Based on these results, we recommend including a minimum of five filters in any SED fitting procedure, with at least two at a wavelength shorter than 5000 Å (necessary to constrain metallicity) and at least one at a wavelength longer than 8000 Å (necessary to constrain mass).  For metallicity fitting specifically, we recommend making use of every available bandpass in the blue portion of the GC spectrum, and we note that metallicity uncertainty will be larger than mass uncertainty given the discrete nature of available model spectra.

\section{Conclusions} \label{sec:conclusions}

We fitted SEDs to 29 of NGC 4874's GCs with high-quality photometry in ten HST filters.  The SED fitting process allowed us to estimate each GC metallicity and mass, and combining the fitted properties with luminosities derived from photometry gave us M/L ratios slightly higher than those of average local GCs but not outside the realm of observed cases.  Adjusting assumptions about GC age and IMF had a minimal effect on metallicity estimates, in line with our expectation that metallicity is the primary driver of variations in GC SED shape.

In summary, the potential for SED fitting for derivation of GC physical properties in remote galaxies is promising.  Given that the main limitation for this methodology is the requirement for photometry at several different wavelengths, and that GCs in nearby galaxies are typically only imaged in two \textit{HST} filters (and almost never the ten available for NGC 4874), followup research should focus on combining HST data with JWST data as it becomes available for GCs in other galaxies; at low redshifts, JWST photometry reaches deep within the flat or declining portion of the GC SED in the near-infrared, which is key for constraining GC mass estimates.  Finally, this methodology can be used for the higher-redshift GCs in JWST archival data; we are currently working on such material.

\begin{acknowledgments}
We acknowledge financial support from the Natural Sciences and Engineering Research Council of Canada (NSERC) through a Discovery Grant to WEH.  KH thanks her clubmates at West End Swords for keeping her caffeinated and exercised while writing.
\end{acknowledgments}

\section*{Data Availability}

The HST data presented in this article were obtained from the Mikulski Archive for Space Telescopes (MAST) at the Space Telescope Science Institute. The specific observations analyzed can be accessed via \dataset[doi: 10.17909/y5vm-sk23]{https://doi.org/10.17909/y5vm-sk23}.

\hfill \\

\begin{contribution}

KH extracted photometry, developed the SED fitting routine, created the figures, and wrote and submitted the manuscript.  WEH led the program from which the images were taken, assisted with photometry, and edited the manuscript.  JK calculated effective wavelengths and provided code for filter removal analysis.

\end{contribution}

%
\facilities{\textit{HST} (ACS, WFC3)}

\software{Python (\href{https://www.python.org}{https://www.python.org}), NumPy \citep{harris2020numpy}, pandas \citep{team2020pandas}, SciPy \citep{virtanen2020scipy}, Matplotlib \citep{hunter2007matplotlib}, DOLPHOT \citep{dolphin2000dolphot}}

\bibliography{Bibliography}{}

\begin{thebibliography}{}
\expandafter\ifx\csname natexlab\endcsname\relax\def\natexlab#1{#1}\fi
\providecommand{\url}[1]{\href{#1}{#1}}
\providecommand{\dodoi}[1]{doi:~\href{http://doi.org/#1}{\nolinkurl{#1}}}
\providecommand{\doeprint}[1]{\href{http://ascl.net/#1}{\nolinkurl{http://ascl.net/#1}}}
\providecommand{\doarXiv}[1]{\href{https://arxiv.org/abs/#1}{\nolinkurl{https://arxiv.org/abs/#1}}}

\bibitem[{H. {Baumgardt} {et~al.}(2020){Baumgardt}, {Sollima}, \&
  {Hilker}}]{baumgardt2020masstolight}
{Baumgardt}, H., {Sollima}, A., \& {Hilker}, M. 2020, \bibinfo{title}{{Absolute
  V-band magnitudes and mass-to-light ratios of Galactic globular clusters},}
  \pasa, 37, e046, \dodoi{10.1017/pasa.2020.38}

\bibitem[{R. {Bezanson} {et~al.}(2024){Bezanson}, {Labbe}, {Whitaker}, {Leja},
  {Price}, {Franx}, {Brammer}, {Marchesini}, {Zitrin}, {Wang}, {Weaver},
  {Furtak}, {Atek}, {Coe}, {Cutler}, {Dayal}, {van Dokkum}, {Feldmann},
  {F{\"o}rster Schreiber}, {Fujimoto}, {Geha}, {Glazebrook}, {de Graaff},
  {Greene}, {Juneau}, {Kassin}, {Kriek}, {Khullar}, {Maseda}, {Mowla},
  {Muzzin}, {Nanayakkara}, {Nelson}, {Oesch}, {Pacifici}, {Pan}, {Papovich},
  {Setton}, {Shapley}, {Smit}, {Stefanon}, {Taylor}, \&
  {Williams}}]{bezanson2024uncover}
{Bezanson}, R., {Labbe}, I., {Whitaker}, K.~E., {et~al.} 2024,
  \bibinfo{title}{{The JWST UNCOVER Treasury Survey: Ultradeep NIRSpec and
  NIRCam Observations before the Epoch of Reionization},} \apj, 974, 92,
  \dodoi{10.3847/1538-4357/ad66cf}

\bibitem[{A. {Calamida} {et~al.}(2022){Calamida}, {Bajaj}, {Mack}, {Marinelli},
  {Medina}, {Pidgeon}, {Kozhurina-Platais}, {Shanahan}, \&
  {Som}}]{calamida2022zerosWFC3}
{Calamida}, A., {Bajaj}, V., {Mack}, J., {et~al.} 2022, \bibinfo{title}{{New
  Photometric Calibration of the Wide Field Camera 3 Detectors},} \aj, 164, 32,
  \dodoi{10.3847/1538-3881/ac73f0}

\bibitem[{G. {Chabrier}(2001){Chabrier}}]{chabrier2001imf}
{Chabrier}, G. 2001, \bibinfo{title}{{The Galactic Disk Mass Budget. I. Stellar
  Mass Function and Density},} \apj, 554, 1274, \dodoi{10.1086/321401}

\bibitem[{A.~E. {Dolphin}(2000){Dolphin}}]{dolphin2000dolphot}
{Dolphin}, A.~E. 2000, \bibinfo{title}{{WFPC2 Stellar Photometry with
  HSTPHOT},} \pasp, 112, 1383, \dodoi{10.1086/316630}

\bibitem[{A. {Dumont} {et~al.}(2022){Dumont}, {Seth}, {Strader}, {Voggel},
  {Sand}, {Hughes}, {Caldwell}, {Crnojevi{\'c}}, {Mateo}, {Bailey}, \&
  {Forbes}}]{dumont2022masstolight}
{Dumont}, A., {Seth}, A.~C., {Strader}, J., {et~al.} 2022, \bibinfo{title}{{A
  Population of Luminous Globular Clusters and Stripped Nuclei with Elevated
  Mass to Light Ratios around NGC 5128},} \apj, 929, 147,
  \dodoi{10.3847/1538-4357/ac551c}

\bibitem[{K. {Fahrion} {et~al.}(2020){Fahrion}, {Lyubenova}, {Hilker}, {van de
  Ven}, {Falc{\'o}n-Barroso}, {Leaman}, {Mart{\'\i}n-Navarro}, {Bittner},
  {Coccato}, {Corsini}, {Gadotti}, {Iodice}, {McDermid}, {Pinna}, {Sarzi},
  {Viaene}, {de Zeeuw}, \& {Zhu}}]{fahrion2020spectroscopy}
{Fahrion}, K., {Lyubenova}, M., {Hilker}, M., {et~al.} 2020,
  \bibinfo{title}{{The Fornax 3D project: Non-linear colour-metallicity
  relation of globular clusters},} \aap, 637, A27,
  \dodoi{10.1051/0004-6361/202037686}

\bibitem[{A.~L. {Faisst} {et~al.}(2022){Faisst}, {Chary}, {Brammer}, \&
  {Toft}}]{faisst2022tinythings}
{Faisst}, A.~L., {Chary}, R.~R., {Brammer}, G., \& {Toft}, S. 2022,
  \bibinfo{title}{{What Are Those Tiny Things? A First Study of Compact Star
  Clusters in the SMACS0723 Field with JWST},} \apjl, 941, L11,
  \dodoi{10.3847/2041-8213/aca1bf}

\bibitem[{Z. {Fan} {et~al.}(2016){Fan}, {de Grijs}, {Chen}, {Jiang}, {Bian}, \&
  {Li}}]{fan2016m31sed}
{Fan}, Z., {de Grijs}, R., {Chen}, B., {et~al.} 2016, \bibinfo{title}{{Lick
  Indices and Spectral Energy Distribution Analysis Based on an M31 Star
  Cluster Sample: Comparisons of Methods and Models},} \aj, 152, 208,
  \dodoi{10.3847/0004-6256/152/6/208}

\bibitem[{Z. {Fan} {et~al.}(2018){Fan}, {Li}, \& {Zhao}}]{fan2018m31sed}
{Fan}, Z., {Li}, Z., \& {Zhao}, G. 2018, \bibinfo{title}{{The Ages of M31 Star
  Clusters: Spectral Energy Distribution versus Color-Magnitude Diagram},} \aj,
  156, 191, \dodoi{10.3847/1538-3881/aae1aa}

\bibitem[{S. {Fujimoto} {et~al.}(2025){Fujimoto}, {Coe}, {Abdurro'uf},
  {Abraham}, {Adamo}, {Akins}, {Amorin}, {Arrabal Haro}, {Asada}, {Atek},
  {Bagley}, {Bhatawdekar}, {Bradac}, {Bradley}, {Brammer}, {Bromm}, {Casey},
  {Chisholm}, {Conselice}, {Dai}, {Dayal}, {Desprez}, {Dessauges-Zavadsky},
  {Dickinson}, {Diego}, {Egami}, {Eisenstein}, {Ferguson}, {Finkelstein},
  {Furtak}, {Hamilton}, {Harikane}, {Hashimoto}, {Hathi}, {Hsiao}, {Inayoshi},
  {Jimenez-Teja}, {Jogee}, {Kartaltepe}, {Koekemoer}, {Kohno}, {Kokorev},
  {Kumari}, {Labbe}, {Larson}, {Lucas}, {Magdis}, {Marchesini}, {Markov},
  {Martis}, {Matthee}, {Meena}, {Messa}, {Mowla}, {Munoz}, {Naidu}, {Nakajima},
  {Nakane}, {Noirot}, {Oesch}, {Oguri}, {Ono}, {Ouchi}, {Pan}, {Papovich},
  {Pascale}, {Pierel}, {Postman}, {Resseguier}, {Rest}, {Richard}, {Ricotti},
  {Rigby}, {Sawicki}, {Schneider}, {Shimasaku}, {Strolger}, {Sun}, {Toft},
  {Tripodi}, {Trussler}, {Tsujita}, {Umeda}, {Valentino}, {Vanzella},
  {Venditti}, {Watson}, {Weaver}, {Welch}, {Willott}, {Windhorst}, {Xu},
  {Yanagisawa}, {Zackrisson}, \& {Zitrin}}]{fujimoto2025venus}
{Fujimoto}, S., {Coe}, D., {Abdurro'uf}, A., {et~al.} 2025, {Vast Exploration
  for Nascent, Unexplored Sources (VENUS)},, JWST Proposal. Cycle 4, ID. \#6882

\bibitem[{L. {Girardi} {et~al.}(2000){Girardi}, {Bressan}, {Bertelli}, \&
  {Chiosi}}]{girardi2000padova}
{Girardi}, L., {Bressan}, A., {Bertelli}, G., \& {Chiosi}, C. 2000,
  \bibinfo{title}{{Evolutionary tracks and isochrones for low- and
  intermediate-mass stars: From 0.15 to 7 M$_{sun}$, and from Z=0.0004 to
  0.03},} \aaps, 141, 371, \dodoi{10.1051/aas:2000126}

\bibitem[{C.~R. Harris {et~al.}(2020)Harris, Millman, van~der Walt, Gommers,
  Virtanen, Cournapeau, Wieser, Taylor, Berg, Smith, Kern, Picus, Hoyer, van
  Kerkwijk, Brett, Haldane, del R{\'{i}}o, Wiebe, Peterson,
  G{\'{e}}rard-Marchant, Sheppard, Reddy, Weckesser, Abbasi, Gohlke, \&
  Oliphant}]{harris2020numpy}
Harris, C.~R., Millman, K.~J., van~der Walt, S.~J., {et~al.} 2020,
  \bibinfo{title}{Array programming with {NumPy},} Nature, 585, 357,
  \dodoi{10.1038/s41586-020-2649-2}

\bibitem[{W.~E. {Harris}(1996){Harris}}]{harris1996catalogue}
{Harris}, W.~E. 1996, \bibinfo{title}{{A Catalog of Parameters for Globular
  Clusters in the Milky Way},} \aj, 112, 1487, \dodoi{10.1086/118116}

\bibitem[{W.~E. {Harris}(2010){Harris}}]{harris2010catalogue}
{Harris}, W.~E. 2010, \bibinfo{title}{{A New Catalog of Globular Clusters in
  the Milky Way},} arXiv e-prints, arXiv:1012.3224,
  \dodoi{10.48550/arXiv.1012.3224}

\bibitem[{W.~E. {Harris}(2023){Harris}}]{harris2023colours}
{Harris}, W.~E. 2023, \bibinfo{title}{{A Photometric Survey of Globular Cluster
  Systems in Brightest Cluster Galaxies},} \apjs, 265, 9,
  \dodoi{10.3847/1538-4365/acab5c}

\bibitem[{W.~E. {Harris} {et~al.}(2017{\natexlab{a}}){Harris}, {Blakeslee}, \&
  {Harris}}]{harris2017masstolight}
{Harris}, W.~E., {Blakeslee}, J.~P., \& {Harris}, G. L.~H. 2017{\natexlab{a}},
  \bibinfo{title}{{Galactic Dark Matter Halos and Globular Cluster Populations.
  III. Extension to Extreme Environments},} \apj, 836, 67,
  \dodoi{10.3847/1538-4357/836/1/67}

\bibitem[{W.~E. {Harris} {et~al.}(2017{\natexlab{b}}){Harris}, {Ciccone},
  {Eadie}, {Gnedin}, {Geisler}, {Rothberg}, \& {Bailin}}]{harris2017BCGs}
{Harris}, W.~E., {Ciccone}, S.~M., {Eadie}, G.~M., {et~al.} 2017{\natexlab{b}},
  \bibinfo{title}{{Globular Cluster Systems in Brightest Cluster Galaxies. III:
  Beyond Bimodality},} \apj, 835, 101, \dodoi{10.3847/1538-4357/835/1/101}

\bibitem[{W.~E. {Harris} {et~al.}(2013){Harris}, {Harris}, \&
  {Alessi}}]{harris2013stellarmass}
{Harris}, W.~E., {Harris}, G. L.~H., \& {Alessi}, M. 2013, \bibinfo{title}{{A
  Catalog of Globular Cluster Systems: What Determines the Size of a Galaxy's
  Globular Cluster Population?},} \apj, 772, 82,
  \dodoi{10.1088/0004-637X/772/2/82}

\bibitem[{W.~E. {Harris} \& M. {Reina-Campos}(2023){Harris} \&
  {Reina-Campos}}]{harris2023jwst}
{Harris}, W.~E., \& {Reina-Campos}, M. 2023, \bibinfo{title}{{JWST photometry
  of globular cluster populations in Abell 2744 at z = 0.3},} \mnras, 526,
  2696, \dodoi{10.1093/mnras/stad2903}

\bibitem[{W.~E. {Harris} \& M. {Reina-Campos}(2024){Harris} \&
  {Reina-Campos}}]{harris2024jwst}
{Harris}, W.~E., \& {Reina-Campos}, M. 2024, \bibinfo{title}{{JWST Photometry
  of Globular Clusters in A2744. II. Luminosity and Color Distributions},}
  \apj, 971, 155, \dodoi{10.3847/1538-4357/ad583c}

\bibitem[{W.~E. {Harris} {et~al.}(2014){Harris}, {Morningstar}, {Gnedin},
  {O'Halloran}, {Blakeslee}, {Whitmore}, {C{\^o}t{\'e}}, {Geisler}, {Peng},
  {Bailin}, {Rothberg}, {Cockcroft}, \& {Barber DeGraaff}}]{harris2014gclum}
{Harris}, W.~E., {Morningstar}, W., {Gnedin}, O.~Y., {et~al.} 2014,
  \bibinfo{title}{{Globular Cluster Systems in Brightest Cluster Galaxies: A
  Near-universal Luminosity Function?},} \apj, 797, 128,
  \dodoi{10.1088/0004-637X/797/2/128}

\bibitem[{K. {Hartman} \& W.~E. {Harris}(2024){Hartman} \&
  {Harris}}]{hartman2024models}
{Hartman}, K., \& {Harris}, W.~E. 2024, \bibinfo{title}{{The Effect of Age and
  Stellar Model Choice on Globular Cluster Color-to-metallicity Conversions},}
  \aj, 168, 75, \dodoi{10.3847/1538-3881/ad57bc}

\bibitem[{K. {Hartman} {et~al.}(2023){Hartman}, {Harris}, {Blakeslee}, {Ma}, \&
  {Greene}}]{hartman2023enviro}
{Hartman}, K., {Harris}, W.~E., {Blakeslee}, J.~P., {Ma}, C.-P., \& {Greene},
  J.~E. 2023, \bibinfo{title}{{Comparing Globular Cluster System Properties
  with Host Galaxy Environment},} \apj, 953, 154,
  \dodoi{10.3847/1538-4357/ace340}

\bibitem[{J.~D. {Hunter}(2007){Hunter}}]{hunter2007matplotlib}
{Hunter}, J.~D. 2007, \bibinfo{title}{{Matplotlib: A 2D Graphics Environment},}
  Computing in Science and Engineering, 9, 90, \dodoi{10.1109/MCSE.2007.55}

\bibitem[{P. {Kroupa} {et~al.}(1993){Kroupa}, {Tout}, \&
  {Gilmore}}]{kroupa1993imf}
{Kroupa}, P., {Tout}, C.~A., \& {Gilmore}, G. 1993, \bibinfo{title}{{The
  Distribution of Low-Mass Stars in the Galactic Disc},} \mnras, 262, 545,
  \dodoi{10.1093/mnras/262.3.545}

\bibitem[{S.~S. {Larsen} {et~al.}(2022){Larsen}, {Eitner}, {Magg}, {Bergemann},
  {Moltzer}, {Brodie}, {Romanowsky}, \& {Strader}}]{larsen2022metallicities}
{Larsen}, S.~S., {Eitner}, P., {Magg}, E., {et~al.} 2022, \bibinfo{title}{{The
  chemical composition of globular clusters in the Local Group},} \aap, 660,
  A88, \dodoi{10.1051/0004-6361/202142243}

\bibitem[{T. pandas~development team(2020)pandas~development
  team}]{team2020pandas}
pandas~development team, T. 2020, pandas-dev/pandas: Pandas, latest Zenodo,
  \dodoi{10.5281/zenodo.3509134}

\bibitem[{N. {Pastorello} {et~al.}(2015){Pastorello}, {Forbes}, {Usher},
  {Brodie}, {Romanowsky}, {Strader}, {Spitler}, {Alabi}, {Foster}, {Jennings},
  {Kartha}, \& {Pota}}]{pastorello2015met}
{Pastorello}, N., {Forbes}, D.~A., {Usher}, C., {et~al.} 2015,
  \bibinfo{title}{{The SLUGGS survey: combining stellar and globular cluster
  metallicities in the outer regions of early-type galaxies},} \mnras, 451,
  2625, \dodoi{10.1093/mnras/stv1131}

\bibitem[{E.~W. {Peng} {et~al.}(2006){Peng}, {Jord{\'a}n}, {C{\^o}t{\'e}},
  {Blakeslee}, {Ferrarese}, {Mei}, {West}, {Merritt}, {Milosavljevi{\'c}}, \&
  {Tonry}}]{peng2006nonlinear}
{Peng}, E.~W., {Jord{\'a}n}, A., {C{\^o}t{\'e}}, P., {et~al.} 2006,
  \bibinfo{title}{{The ACS Virgo Cluster Survey. IX. The Color Distributions of
  Globular Cluster Systems in Early-Type Galaxies},} \apj, 639, 95,
  \dodoi{10.1086/498210}

\bibitem[{E.~W. {Peng} {et~al.}(2011){Peng}, {Ferguson}, {Goudfrooij},
  {Hammer}, {Lucey}, {Marzke}, {Puzia}, {Carter}, {Balcells}, {Bridges},
  {Chiboucas}, {del Burgo}, {Graham}, {Guzm{\'a}n}, {Hudson}, {Matkovi{\'c}},
  {Merritt}, {Miller}, {Mouhcine}, {Phillipps}, {Sharples}, {Smith}, {Tully},
  \& {Verdoes Kleijn}}]{peng2011richGCS}
{Peng}, E.~W., {Ferguson}, H.~C., {Goudfrooij}, P., {et~al.} 2011,
  \bibinfo{title}{{The HST/ACS Coma Cluster Survey. IV. Intergalactic Globular
  Clusters and the Massive Globular Cluster System at the Core of the Coma
  Galaxy Cluster},} \apj, 730, 23, \dodoi{10.1088/0004-637X/730/1/23}

\bibitem[{A. {Pietrinferni} {et~al.}(2004){Pietrinferni}, {Cassisi}, {Salaris},
  \& {Castelli}}]{pietrinferni2004basti}
{Pietrinferni}, A., {Cassisi}, S., {Salaris}, M., \& {Castelli}, F. 2004,
  \bibinfo{title}{{A Large Stellar Evolution Database for Population Synthesis
  Studies. I. Scaled Solar Models and Isochrones},} \apj, 612, 168,
  \dodoi{10.1086/422498}

\bibitem[{M. {Reina-Campos} \& W.~E. {Harris}(2024){Reina-Campos} \&
  {Harris}}]{reinacampos2024rescuer}
{Reina-Campos}, M., \& {Harris}, W.~E. 2024, \bibinfo{title}{{RESCUER:
  cosmological K-corrections for star clusters},} \mnras, 531, 4099,
  \dodoi{10.1093/mnras/stae1414}

\bibitem[{E.~E. {Salpeter}(1955){Salpeter}}]{salpeter1955imf}
{Salpeter}, E.~E. 1955, \bibinfo{title}{{The Luminosity Function and Stellar
  Evolution.},} \apj, 121, 161, \dodoi{10.1086/145971}

\bibitem[{E.~F. {Schlafly} \& D.~P. {Finkbeiner}(2011){Schlafly} \&
  {Finkbeiner}}]{schlafly2011reddening}
{Schlafly}, E.~F., \& {Finkbeiner}, D.~P. 2011, \bibinfo{title}{{Measuring
  Reddening with Sloan Digital Sky Survey Stellar Spectra and Recalibrating
  SFD},} \apj, 737, 103, \dodoi{10.1088/0004-637X/737/2/103}

\bibitem[{B. {Sinnott} {et~al.}(2010){Sinnott}, {Hou}, {Anderson}, {Harris}, \&
  {Woodley}}]{sinnott2010colour}
{Sinnott}, B., {Hou}, A., {Anderson}, R., {Harris}, W.~E., \& {Woodley}, K.~A.
  2010, \bibinfo{title}{{New g'r'i'z' Photometry of the NGC 5128 Globular
  Cluster System},} \aj, 140, 2101, \dodoi{10.1088/0004-6256/140/6/2101}

\bibitem[{J. {Strader} {et~al.}(2011){Strader}, {Caldwell}, \&
  {Seth}}]{strader2011m31}
{Strader}, J., {Caldwell}, N., \& {Seth}, A.~C. 2011, \bibinfo{title}{{Star
  Clusters in M31. V. Internal Dynamical Trends: Some Troublesome, Some
  Reassuring},} \aj, 142, 8, \dodoi{10.1088/0004-6256/142/1/8}

\bibitem[{A.~T. {Tokunaga} \& W.~D. {Vacca}(2005){Tokunaga} \&
  {Vacca}}]{tokunaga2005effwave}
{Tokunaga}, A.~T., \& {Vacca}, W.~D. 2005, \bibinfo{title}{{The Mauna Kea
  Observatories Near-Infrared Filter Set. III. Isophotal Wavelengths and
  Absolute Calibration},} \pasp, 117, 421, \dodoi{10.1086/429382}

\bibitem[{C. {Usher} {et~al.}(2012){Usher}, {Forbes}, {Brodie}, {Foster},
  {Spitler}, {Arnold}, {Romanowsky}, {Strader}, \& {Pota}}]{usher2012colour}
{Usher}, C., {Forbes}, D.~A., {Brodie}, J.~P., {et~al.} 2012,
  \bibinfo{title}{{The SLUGGS survey: calcium triplet-based spectroscopic
  metallicities for over 900 globular clusters},} \mnras, 426, 1475,
  \dodoi{10.1111/j.1365-2966.2012.21801.x}

\bibitem[{A. {Vazdekis} {et~al.}(2016){Vazdekis}, {Koleva}, {Ricciardelli},
  {R{\"o}ck}, \& {Falc{\'o}n-Barroso}}]{vazdekis2016EMILES}
{Vazdekis}, A., {Koleva}, M., {Ricciardelli}, E., {R{\"o}ck}, B., \&
  {Falc{\'o}n-Barroso}, J. 2016, \bibinfo{title}{{UV-extended E-MILES stellar
  population models: young components in massive early-type galaxies},} \mnras,
  463, 3409, \dodoi{10.1093/mnras/stw2231}

\bibitem[{D. {Villegas} {et~al.}(2010){Villegas}, {Jord{\'a}n}, {Peng},
  {Blakeslee}, {C{\^o}t{\'e}}, {Ferrarese}, {Kissler-Patig}, {Mei}, {Infante},
  {Tonry}, \& {West}}]{villegas2010gclum}
{Villegas}, D., {Jord{\'a}n}, A., {Peng}, E.~W., {et~al.} 2010,
  \bibinfo{title}{{The ACS Fornax Cluster Survey. VIII. The Luminosity Function
  of Globular Clusters in Virgo and Fornax Early-type Galaxies and Its Use as a
  Distance Indicator},} \apj, 717, 603, \dodoi{10.1088/0004-637X/717/2/603}

\bibitem[{P. Virtanen {et~al.}(2020)Virtanen, Gommers, Oliphant, Haberland,
  Reddy, Cournapeau, Burovski, Peterson, Weckesser, Bright, {van der Walt},
  Brett, Wilson, Millman, Mayorov, Nelson, Jones, Kern, Larson, Carey, Polat,
  Feng, Moore, {VanderPlas}, Laxalde, Perktold, Cimrman, Henriksen, Quintero,
  Harris, Archibald, Ribeiro, Pedregosa, {van Mulbregt}, \& {SciPy 1.0
  Contributors}}]{virtanen2020scipy}
Virtanen, P., Gommers, R., Oliphant, T.~E., {et~al.} 2020,
  \bibinfo{title}{{{SciPy} 1.0: Fundamental Algorithms for Scientific Computing
  in Python},} Nature Methods, 17, 261, \dodoi{10.1038/s41592-019-0686-2}

\bibitem[{C.~N.~A. {Willmer}(2018){Willmer}}]{willmer2018solarmags}
{Willmer}, C. N.~A. 2018, \bibinfo{title}{{The Absolute Magnitude of the Sun in
  Several Filters},} \apjs, 236, 47, \dodoi{10.3847/1538-4365/aabfdf}

\bibitem[{C.~J. {Willott} {et~al.}(2022){Willott}, {Doyon}, {Albert},
  {Brammer}, {Dixon}, {Muzic}, {Ravindranath}, {Scholz}, {Abraham}, {Artigau},
  {Brada{\v{c}}}, {Goudfrooij}, {Hutchings}, {Iyer}, {Jayawardhana}, {LaMassa},
  {Martis}, {Meyer}, {Morishita}, {Mowla}, {Muzzin}, {Noirot}, {Pacifici},
  {Rowlands}, {Sarrouh}, {Sawicki}, {Taylor}, {Volk}, \&
  {Zabl}}]{willott2022canucs}
{Willott}, C.~J., {Doyon}, R., {Albert}, L., {et~al.} 2022,
  \bibinfo{title}{{The Near-infrared Imager and Slitless Spectrograph for the
  James Webb Space Telescope. II. Wide Field Slitless Spectroscopy},} \pasp,
  134, 025002, \dodoi{10.1088/1538-3873/ac5158}

\bibitem[{R.~A. {Windhorst} {et~al.}(2023){Windhorst}, {Cohen}, {Jansen},
  {Summers}, {Tompkins}, {Conselice}, {Driver}, {Yan}, {Coe}, {Frye}, {Grogin},
  {Koekemoer}, {Marshall}, {O'Brien}, {Pirzkal}, {Robotham}, {Ryan}, {Willmer},
  {Carleton}, {Diego}, {Keel}, {Porto}, {Redshaw}, {Scheller}, {Wilkins},
  {Willner}, {Zitrin}, {Adams}, {Austin}, {Arendt}, {Beacom}, {Bhatawdekar},
  {Bradley}, {Broadhurst}, {Cheng}, {Civano}, {Dai}, {Dole}, {D'Silva},
  {Duncan}, {Fazio}, {Ferrami}, {Ferreira}, {Finkelstein}, {Furtak}, {Gim},
  {Griffiths}, {Hammel}, {Harrington}, {Hathi}, {Holwerda}, {Honor}, {Huang},
  {Hyun}, {Im}, {Joshi}, {Kamieneski}, {Kelly}, {Larson}, {Li}, {Lim}, {Ma},
  {Maksym}, {Manzoni}, {Meena}, {Milam}, {Nonino}, {Pascale}, {Petric},
  {Pierel}, {Polletta}, {R{\"o}ttgering}, {Rutkowski}, {Smail}, {Straughn},
  {Strolger}, {Swirbul}, {Trussler}, {Wang}, {Welch}, {B. Wyithe}, {Yun},
  {Zackrisson}, {Zhang}, \& {Zhao}}]{windhorst2023pearls}
{Windhorst}, R.~A., {Cohen}, S.~H., {Jansen}, R.~A., {et~al.} 2023,
  \bibinfo{title}{{JWST PEARLS. Prime Extragalactic Areas for Reionization and
  Lensing Science: Project Overview and First Results},} \aj, 165, 13,
  \dodoi{10.3847/1538-3881/aca163}

\bibitem[{K.~A. {Woodley} {et~al.}(2010){Woodley}, {Harris}, {Puzia},
  {G{\'o}mez}, {Harris}, \& {Geisler}}]{woodley2010agemet}
{Woodley}, K.~A., {Harris}, W.~E., {Puzia}, T.~H., {et~al.} 2010,
  \bibinfo{title}{{The Ages, Metallicities, and Alpha Element Enhancements of
  Globular Clusters in the Elliptical NGC 5128: A Homogeneous Spectroscopic
  Study with Gemini/Gemini Multi-Object Spectrograph},} \apj, 708, 1335,
  \dodoi{10.1088/0004-637X/708/2/1335}

\end{thebibliography}
\bibliographystyle{aasjournalv7}

\end{document}